%% file: main.tex
\documentclass[Afour,sageh,review]{sagej}

\input{samiyacommands}

\usepackage{moreverb,url}

\usepackage{tikz}
\usetikzlibrary{positioning}

\newcommand{\removeInShortVer}[1]{#1}

\usepackage{perpage}
\MakePerPage{footnote}

\usepackage[colorlinks,bookmarksopen,bookmarksnumbered,citecolor=red,urlcolor=red]{hyperref}

\newcommand\BibTeX{{\rmfamily B\kern-.05em \textsc{i\kern-.025em b}\kern-.08em
T\kern-.1667em\lower.7ex\hbox{E}\kern-.125emX}}

\def\volumeyear{2025}

\begin{document}

\runninghead{Alkhairy 2025}

\title{Auditory Filter Behavior and Updated Estimated Constants}

\author{Samiya A Alkhairy\affilnum{1}}

\affiliation{\affilnum{1}Massachusetts Institute of Technology, USA and King Abdulaziz City for Science and Technology, KSA} 

\corrauth{Samiya Alkhairy}

\email{samiya@mit.edu, samiya@alum.mit.edu}

\begin{abstract}

Filters from the Gammatone family are often used to model auditory signal processing, but the filter constant values used to mimic human hearing are largely set to values based on historical psychoacoustic data collected several decades ago. Here, we move away from this long-standing convention, and estimate filter constants using a range of more recent reported filter characteristics (such as quality factors and ratios between quality factors and peak group delay) within a characteristics-based framework that clarifies how filter behavior is related to the underlying constants.
Using a sharp-filter approximation that captures shared peak-region behavior across certain classes of filters, we analyze the range of behaviors accessible when the full degrees of freedom of the filter are utilized rather than fixing the filter order or exponent to historically prescribed values. Filter behavior is characterized using magnitude-based and phase-based characteristics and their ratios, which reveal which characteristics are informative for constraining filter constants and which are only weakly constraining. We show that these insights and estimation methods extend to multiple realizable filter classes from the Gammatone family and apply them, together with recent physiological and psychoacoustic observations, to derive constraints on and estimates for filter constants for human auditory filters. More broadly, this framework supports the design of auditory filters with arbitrary characteristic-level specifications and enables systematic assessment of how variations in filter characteristics influence auditory models, perceptual findings, and technologies that rely on auditory filterbanks.

\end{abstract}

\keywords{auditory filterbanks, auditory filters, 
ERB, filter design, psychoacoustics, simultaneous masking, forward masking, tuning curves, filter characteristics, group delay, tuning ratio, parameter space, sensitivity, cochlear modeling}

\maketitle


\section{Introduction}

\subsection{Motivation}
Gammatone filters and related classes of filters have typically been used to mimic auditory signal processing. The Gammatone filters used to mimic signal processing in humans have been fixed to fourth order, with a second parameter that is typically set based on equivalent rectangular bandwidths (ERBs) as described in \cite{slaney1993efficient}. The chosen value of the filter order is due to fits to tuning curves from simultaneous masking psychoacoustic measurements that were performed several decades ago by \cite{glasberg1990derivation}.

This leaves us with two questions. The first is regarding understanding the full range of behavior accessible by these filters - beyond fixed order filters, and their implications on the values of filter characteristics such as peak frequencies, 3 dB quality factors, ERBs, and group delays at the peak. The second is regarding how we may design the auditory filters to account for recent information that contradicts the simultaneous masking results. 

Answering the first question allows to determine whether certain types of behavior may be fulfilled by auditory filters, how these map onto requirements on filter characteristics, and the sensitivity of filter characteristics to changes in filter constants over a reasonable range of values. We propose that this analysis is at least as important - if not more important, than constructing new classes of auditory filters and studying the difference between them. 

Addressing the second question requires appropriate methods to design the filters and estimate the filter constants given tuning curves or reported values of filter characteristics. Here we choose the latter, and note that proper methods to design filter given filter characteristics must build on an analysis of the filter behavior as well as the sensitivity of the filter characteristics to filter constants. The alternative approach is to fit the filters to newly reported tuning curves, but this would again result in fixed values of filter constants without an understanding of how they are related to descriptions of filter behavior encoded in reported filter characteristics.

\subsection{Goal and Objective}
These two questions lead us to our goal which is to design human auditory filters given recently reported values of filter characteristics in a manner that allows us to understand how varying the filter characteristics influences the estimated filter constants. Our goal is achieved by fulfilling three objectives: (1) analyzing the behavior of filters and the dependence of filter characteristics on filter constants, (2) improving methods for estimating the filter constants from specifications on sets of filter characteristics, and (3) estimating constraints on and values for the filter constants using recent observations and reported values for filter characteristics relevant for humans.

Our methods and estimates apply to several classes of filters from the Gammatone family. In the next section, we fulfill the first two objectives using a sharp-filter approximation for various filter classes, and in the following section, we demonstrate that our findings and methods for at least three classes of filters that are well-approximated by the sharp-filter approximation in the peak region for the parameter region of interest.

We then proceed to estimating the constraints and filter constant values for our third objective where we exemplify the use of the aforementioned  characteristics-based filter design methods to estimate the values of filter constants for human auditory filterbanks given a few reported values of characteristics. In particular, we use reported values of the a ratio found to be constant in chinchilla and assumed to be the same across species [\cite{shera2010otoacoustic}] and equivalent rectangular bandwidths reported from recent forward masking psychoacoustic experiments [\cite{oxenham2003estimates}].

\subsection{Significance}

Beyond estimating filter constants for human auditory filters based on recent experiments, we present guidelines for the estimation of filter constants given various sets of reported characteristics. The intuition developed, the filter design methods, and the expressions relating filter characteristics and filter constants further allow (1) designing filters given specification of arbitrary values for the filter characteristics (e.g. personalized filters or to mimic different species), and (2) \textit{systematically} studying the dependence of findings of perceptual studies and models as well as the performance of technologies e.g. [\cite{dietz2011auditory, dimitriadis2010effects}] on filter characteristics.

\section{Sharp-filter Approximation}

In this section we discuss a sharp filter approximation of several realizable classes of filters that may be designed to mimic auditory signal processing in humans. This sharp-filter approximation is not realizable but its simplicity enables the analysis of its behavior and development of methods to estimate the filter constants parameterizing its transfer function from reported (or desired) values of filter characteristics. The behavior and estimation methods carry over the the realizable filters of interest as show later in this paper.

\subsection{Transfer Function}

Here we review the transfer function of the sharp-filter approximation of the various classes of filters. We present the continuous frequency domain expressions in normalized frequency,

\begin{equation}
    \B \triangleq \frac{f}{\CF} \,
\end{equation}

which is in terms of the frequency, $f$, and the characteristic frequency, $\CF$. The time domain expressions are in scaled time, 

\begin{equation}
    \ttilde \triangleq 2\pi\CF t \;.
\end{equation}

The transfer functions are not in the Laplace domain but rather in normalized continuous complex frequency, $s$, where, 

\begin{equation}
    \Im\{s\} = \B \;.
\end{equation}

The transfer functions are of the form,

\begin{equation}
    \HSharp \propto (s-p)^{-\Bu} \;,
    \label{eq:Hsharp}
\end{equation}

where the repeated pole is $p = -\Ap + i\bp$, and the exponent to the base filter transfer function is $\Bu$. We henceforth refer to $(\Ap, \bp, \Bu) \in \mathbb{R}^{3+}$ as the set of filter constants parameterizing the transfer function. 

We note the use of proportionality rather than equality in the above expression as well as in other expressions for transfer functions and impulse responses in this paper. We do so as we are not concerned with mapping the gain constant from the transfer function of one class of filters to its impulse response or to the transfer function of the sharp-filter approximation. We are not concerned with the variation of the gain constants across $\CF$ in a filterbank, or across different classes of filters. When plotting the frequency response, we normalize the magnitude to that at the peak - and reference the phases so that they start at zero.

\begin{figure*}[htbp]
    \centering
    \includegraphics[width=\linewidth]{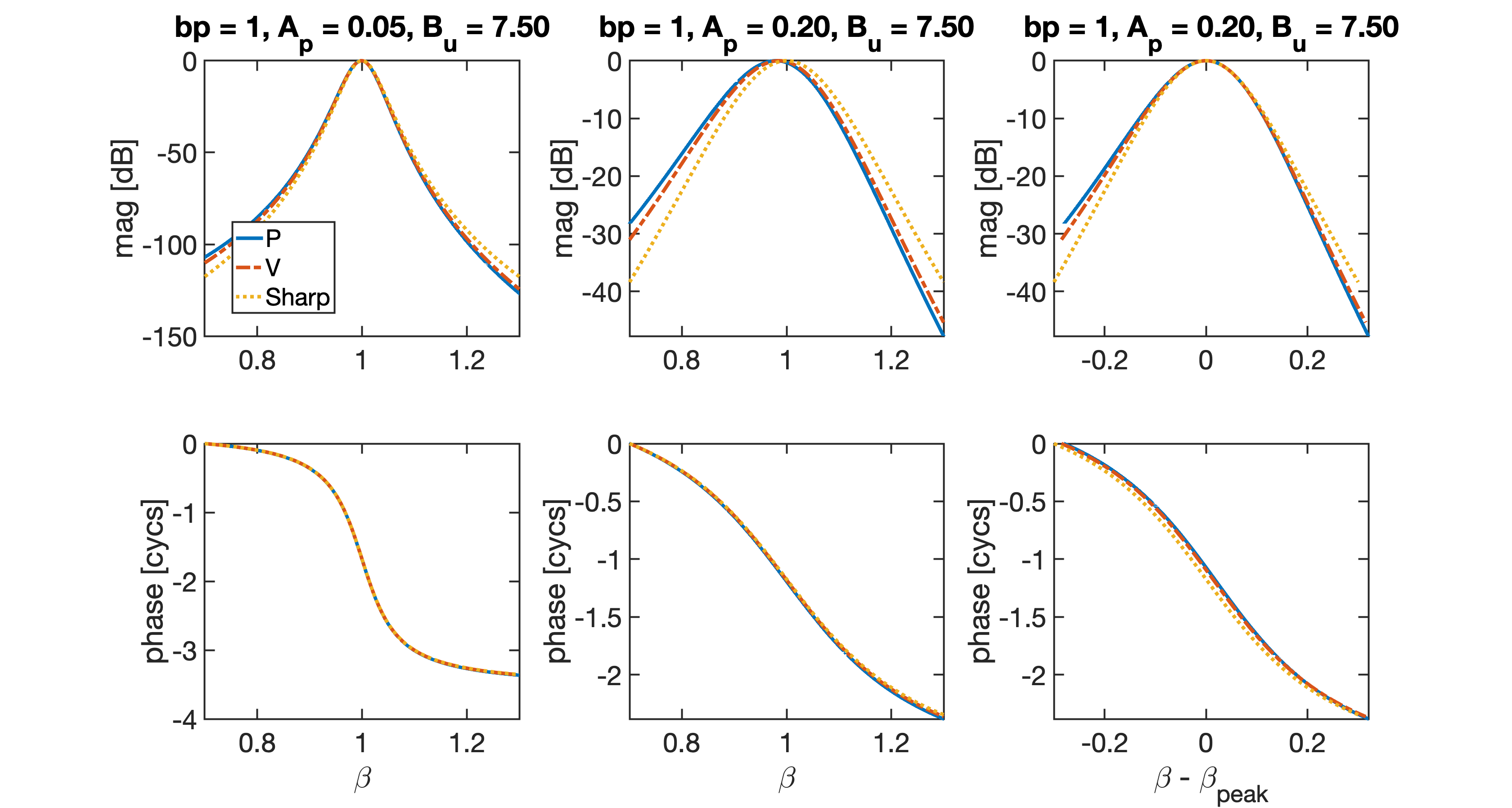}
    \caption[Bode plots]{The left and middle panels show the Bode plots as a function of normalized frequency, $\B$, for the following transfer functions: $\HP$ (blue solid lines), $\HV$ (red dash-dotted lines), and the approximation, $\HSharp$ (yellow dotted lines), which we use to derive our expressions for filter characteristics in terms of filter constants and based upon which the characteristics-based filter design methods are developed. The left and middle panels are generated using two different sets of filter constant values (shown in the panel titles) chosen from the range we deem appropriate in humans. The left panel is more appropriate for filters with larger CFs and the middle panel is for lower CFs. The left and middle shows that $\HSharp$ is a good approximation for the transfer functions for the realizable filters, $\HP$ and $\HV$. In the rightmost panel, we show the validity of this approximation more clearly for the transfer functions from the middle panel by shifting the $x$ axis by $\Bcenter$, which takes on the following values for each of the curves:  $\B_{\textrm{peak}, P} = 0.9798$,  $\B_{\textrm{peak}, V} = 0.9852$, and $\B_{\textrm{peak, Sharp}} = 1$.}
    \label{fig:PandVandSharp}
\end{figure*}

The $\HSharp$ was independently derived as a sharp-filter approximations for both the Generalized-Exponent Filters (GEFs/P) [\cite{alkhairy2024characteristics}] and for the classical Gammatone Filters (GTFs) [\cite{holdsworth1988implementing, darling1991properties}]. We note that the set of filter constants $(\Ap, \bp, \Bu)$ are the native parameters for the transfer functions for GEFs and $V$ (discussed in a later section) and we re-express GTFs in these parameters so as to deal with a single parameterization for all classes of filters. The transfer function $\HSharp$ is particularly good at approximating the aforementioned classes of auditory filters for small values of $\Ap$ and particularly in the region of the peak as shown in figure \ref{fig:PandVandSharp}.

While this class of filters is itself not realizable, the simplicity of its transfer function enables us to derive closed-form expressions for (primarily peak-centric) filter characteristics - e.g. 3dB quality factor, in terms of filter constants. These expressions form the basis upon which we develop our methods to estimate filter constants from reported values of filter characteristics. Over the course of the next few sections, we return to these fundamental expressions, derive expressions for various filter characteristics (including characteristic ratios), study the dependence of the filter characteristics on the pole and filter exponent, and develop accurate methods for the design of filters given values of characteristics, and provide estimates for the filter constants given reported observations and values for filter characteristics.

\subsection{Dependence of Filter Behavior on Constants}

In this section, we show the rich array of possible filter behavior - quantified via filter characteristics, that we can attain provided that we allow ourselves to access the range of values for all degrees of freedom - $\Ap, \Bu, \bp$, rather than restrict ourselves to certain values of $\Bu$ as has typically been the case. We discuss the dependence of filter characteristics on filter constants in order to build intuition about the filters, infer constraints on filter constants given observations, and as the first step towards developing methods to estimate filter constants given specifications on filter characteristics. We perform this analysis here for $\HSharp$ which approximates the common features across realizable classes of filters from the Gammatone family of filters. In a later section, we demonstrate the validity of our findings for realizable filters.

\subsubsection{Expressions for Filter Characteristics}

We limit ourselves to certain frequency domain characteristics and do not include asymmetry of the magnitude of the frequency response which is fundamentally different across the realizable classes of filters in the Gammatone family of filters. We define the filter characteristics in the normalized frequency ($\B$) domain and express them in terms of the set of filter constants  ($\Ap, \bp, \Bu$) that  parameterizes the transfer function. Using the expression for $\HSharp$  (equation \ref{eq:Hsharp}), we derive the following expressions for filter characteristics as functions of the filter constants. These include both magnitude-based characteristics ($\Bcenter, \Qn, \Qerb, \SB$) and phase-based characteristics ($ \B_{maxN}, \NB, \phiaccum$).

\begin{equation}
    \begin{aligned}
        \Bcenter & = \bp \\
        \Qn &  = \frac{\bp}{2\Ap} \bigg( 10^{\frac{n}{10\Bu}} - 1 \bigg)^{-\frac{1}{2}} \\
        \Qerb & = \frac{\bp}{\sqrt{\pi} \Ap} \frac{\Gamma(\Bu)}{\Gamma(\Bu-\frac{1}{2})}\\
        \SB & = \frac{20}{\log(10)} \frac{\Bu}{\Ap^2}\\
        \B_{maxN} & = \bp \\
        \NB & = \frac{1}{2\pi} \frac{\Bu}{\Ap}\\
        \phiaccum & = \frac{\Bu}{2}
    \end{aligned}
    \label{eq:const2chars}
\end{equation}

We derive our expression for $\Qerb$ in the appendix of this paper. The derivation for most of the other characteristics was previously detailed in \cite{alkhairy2024characteristics}.

\subsubsection{Variation of Characteristics with Constants}

In what follows, we describe how the magnitude-based and phase-based filter characteristics of $\HSharp$ vary with the filter constants as may be guided by the closed-form expressions in equation \ref{eq:const2chars}. These descriptions also hold for the realizable classes of auditory filters discussed later.

The first of the magnitude-based characteristics - the peak normalized frequency, $\Bcenter$, of $\HSharp$ depends only on $\bp$ which we typically assign to unity. However, we may choose to vary the value of $\bp$ if we are interested in designing filters for which the peak frequency (best frequency), $f_{peak} = \bp \CF$, is not simply equal to the characteristic frequency. For instance, one may choose to set $\bp < 1$ so that $f_{peak} < \CF$ (e.g. to account for filter behavior at higher SPLs without fully moving to nonlinear filters).

The two quality factors in equation \ref{eq:const2chars} - the quality factor based on an $n$ dB bandwidth, $\Qn$, and the quality factor based on the equivalent rectangular bandwidth (ERB), $\Qerb$ - are both proportional to $\bp$ and inversely proportional to $\Ap$. The quality factors also increase with the filter exponent $\Bu$ but in a more complicated manner. The $n$ dB quality factor is naturally smaller for larger $n$ - e.g. $Q_{15} < Q_3$.

The last magnitude-based filter characteristic in equation \ref{eq:const2chars} is the degree of downward convexity at the peak frequency, $\SB$, in dB. This filter characteristic was introduced in \cite{alkhairy2024characteristics} as the most peak-centric of the magnitude-based filter characteristic as it only depends on the second derivative of the magnitude (in dB) at $\Bcenter$. Similar to the bandwidths that are defined in the $\B$ domain, $\textrm{ERB}_\B$ and $\textrm{BW}_n$, the convexity measure, $\SB$, is a measure of the sharpness of tuning. It is independent of $\bp$ (like ERB$_\B$ and BW$_n$ but unlike $\Qerb$ and $\Qn$), is inversely proportional to $\Ap^2$ and hence highly sensitive to it, and is simply linearly proportional to $\Bu$, rendering it the simplest of the measures for sharpness of tuning. For experimental data, its use is limited by sampling rate and SNR, but it is quite appropriate for models, filters, finely sampled data at high SNR, and fitted data - e.g. \cite{tan2003phenomenological}.

As for the phase-based characteristics: the maximum group delay occurs at $\B_{maxN}$ which in the case of $\HSharp$ occurs at $\Bcenter$. However, in the case of the realizable filters, they are similar to one another but are not exactly equal as may be inferred from figure \ref{fig:PandVandSharp}. 

The maximum group delay or equivalently (for $\HSharp$) the group delay at the peak, $\NB$, in cycles, is linearly proportional to $\Bu$ and inversely proportional to $\Ap$. The phase accumulation, $\phiaccum$ in cycles, is solely dependent on - and directly proportional to, the filter exponent $\Bu$. As expected, if $\Bu = 1$ - i.e. we have a second order filter describing a damped harmonic oscillator, the phase accumulation is half a cycle. 

\begin{figure*}
    \centering
    \includegraphics[width=\linewidth]{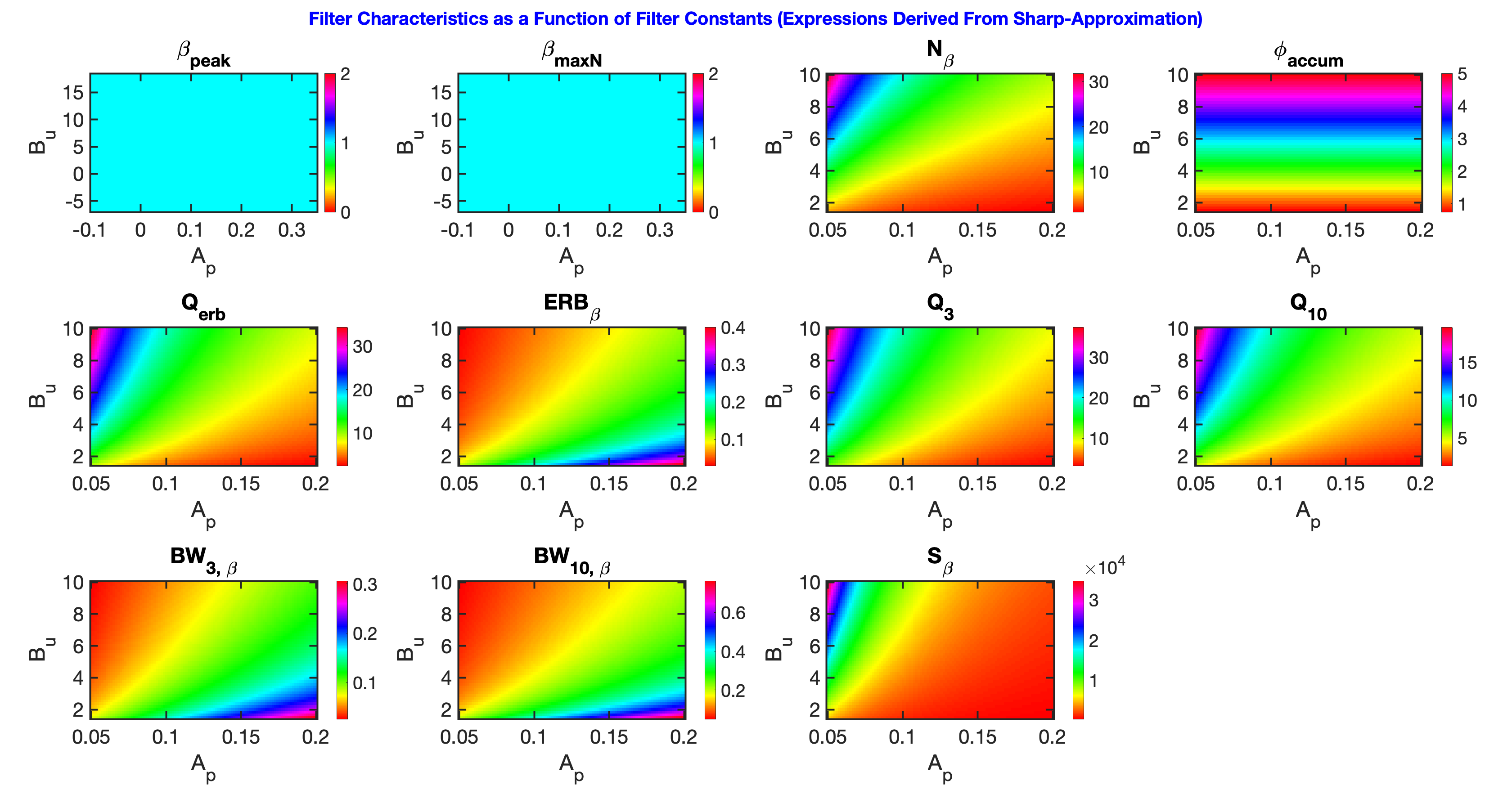}
     \caption[Dependence of filter characteristics on filter constants]{We plot various filter characteristics of $\HSharp$ as a function of filter constants $\Ap$ and $\Bu$ using the closed-form expressions in equation \ref{eq:const2chars}. For the quality factors, $\Bcenter$ and $\beta_{maxN}$, which depend on $\bp$, we used $\bp = 1$ (i.e. $f_{peak} = \CF$). These plots (along with plots of the characteristic ratios in figure \ref{fig:ratios}) may be used to understand the effects of varying the filter exponent and pole on behavior, to understand the relationship between various characteristics, and as a look-up table to quickly estimate filter constants or serve as a stepping stone by providing initial guesses for the characteristics-based filter design methods.}
    \label{fig:chars}
\end{figure*}

These observations regarding the dependence of filter characteristics of $\HSharp$ are visualized in figure \ref{fig:chars}. These plots also may serve as a `look-up' table of sorts for those who design filters by specifying values of filter constants to fulfill certain specifications on filter characteristics. In the figure, we have set $\bp = 1$  (i.e. $f_{peak} = \CF$) for simplicity. We bounded the domain of filter constants to include range - that we find later in the paper, to be most appropriate for filterbanks mimicking auditory signal processing in humans.

\subsubsection{Dependence of Characteristic Ratios on Filter Constants}

One particular type of filter characteristics is defined as combinations of some of the filter characteristics mentioned above. It is relevant to discuss the dependence of these combined characteristics on filter constants as we later used them to estimate filter constants. 

This is motivated by reports from recent work of CF-invariant and/or species-invariant values or patterns in ratios of some of the aforementioned characteristics - e.g. $\frac{\Qerb}{\NB}$ across CF in chinchilla, and $\frac{\Qerb}{Q_{10}}$ across species and CF [\cite{shera2003stimulus, shera2010otoacoustic}]. In what follows, we study the dependence of such characteristic ratios on the filter constants. We limit our interest to characteristic ratios that depend only on $\Bu$ or only on $\Ap$ and which we can derive using expressions in equation \ref{eq:const2chars}.

Of particular interest are $\alpha$ and $g$ which we may express based on the sharp-filter expressions in equation \ref{eq:const2chars} as follows. In a later section, we use reported values for the above characteristic ratios to provide bounds or estimates for filter constants.

\begin{equation}
    \begin{aligned}
        \alpha & \triangleq \frac{\Qerb}{Q_{10}}  = \frac{2}{\sqrt{\pi}} \frac{\Gamma(\Bu)}{\Gamma(\Bu - \frac{1}{2})} \sqrt{10^{1/\Bu} - 1}\\
        g & \triangleq \frac{\Qerb}{\NB}  = 2 \sqrt{\pi} \bp \frac{\Gamma(\Bu)}{\Bu \Gamma(\Bu - \frac{1}{2})}\;.
    \end{aligned}
    \label{eq:charratios}
\end{equation}

\begin{figure}[htbp]
    \subfloat[Ratios($\Bu$)]{\includegraphics[width = 0.48\textwidth]{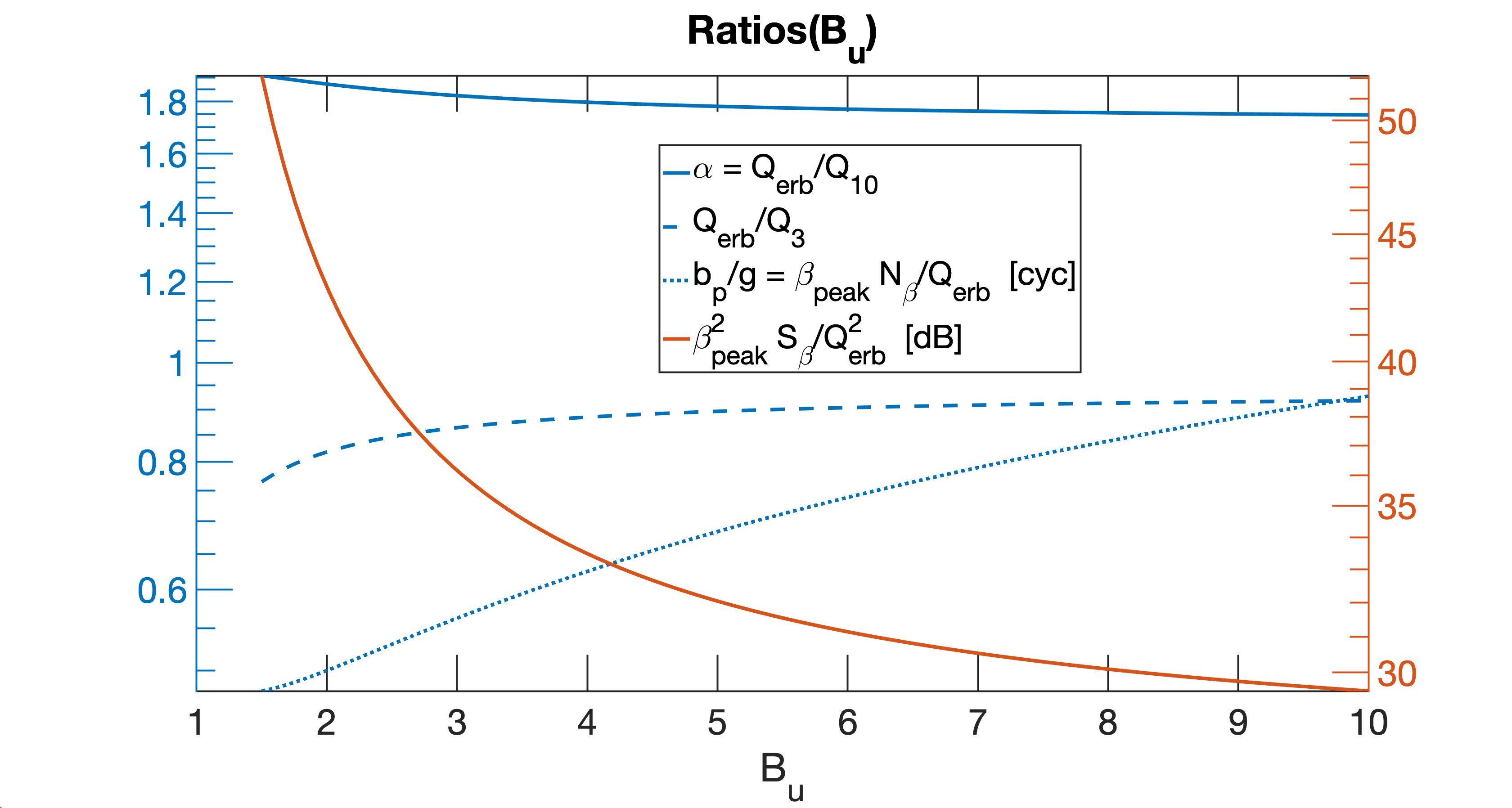} \label{subfig:ratioBu}}
    \\
    \subfloat[Ratios($\Ap$)]{\includegraphics[width = 0.48\textwidth]{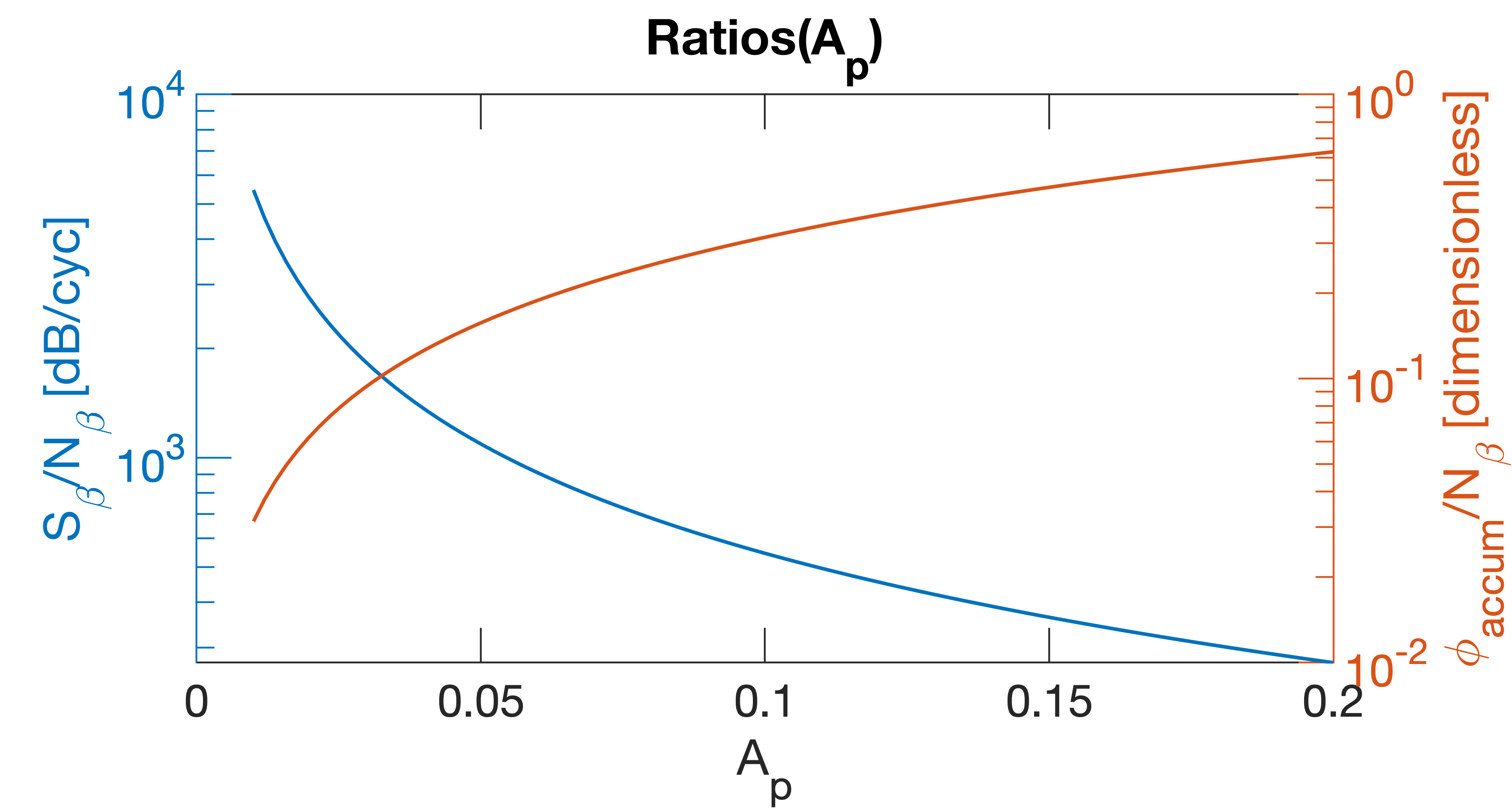} \label{subfig:ratioAp}}
    \caption[Dependence of filter characteristic-ratios on filter constants]{We plot the dependence of several filter characteristics ratios that depend solely on a single filter constant - $\Ap$ or $\Bu$, as derived using the expressions from equation \ref{eq:const2chars} based on $\HSharp$. The top panel reveals that certain $\Bu$-dependent ratios such as $\frac{\Qerb}{Q_{10}}$ are not particularly useful for determining specific values of $\Bu$ due to their slow variation with respect to $\Bu$ in the parameter region of interest. We note that for larger values of $\Bu$, $\frac{g}{\bp}$ and $\frac{\Qerb^2}{\Bcenter^2 \SB}$ also exhibit shallow dependence. In contrast, both ratios that are purely functions of $\Ap$ - i.e. $\frac{\SB}{\NB}$ and $\frac{\phiaccum}{\NB}$, exhibit a strong dependence on $\Ap$ and hence - when available, may reliably be used to determine this filter constant. }
    \label{fig:ratios}
\end{figure}

 In figure \ref{subfig:ratioBu}, we plot a subset of the following ratios of filter characteristics for $\HSharp$ that may be expressed (when using equation \ref{eq:const2chars}) purely as functions of the filter exponent $\Bu$: 
 
 \begin{itemize}
     \item $\frac{\Qerb}{Q_{n}} \equiv \frac{\BWndBbeta}{\ERBB}, \frac{Q_{n}}{Q_{m}} \equiv \frac{\textrm{BW}_{m, \B}}{\textrm{BW}_{n, \B}}$ which are dimensionless
     \item $\frac{\Qerb^2}{\Bcenter^2 \SB} \equiv (\ERBB^2 \SB)^{-1} , \frac{Q_n^2}{\Bcenter^2 \SB} \equiv (\BWndBbeta^2 \SB)^{-1}$ in dB$^{-1}$
     \item $\frac{\Qerb}{\Bcenter \NB} \equiv (\ERBB \NB)^{-1}, \frac{Q_n}{\Bcenter \NB} \equiv (\BWndBbeta \NB)^{-1}$ in cycs$^{-1}$
 \end{itemize}

Reported values for ratios that strongly depend on $\Bu$ may be used to estimate $\Bu$. However, ratios that weakly vary with $\Bu$ over the range of interest such as $\alpha$ cannot be used to determine $\Bu$ - but can  be used to estimate bounds for $\Bu$ with some degree of confidence (as done later in this paper). The same is true for $\frac{g}{\bp}$ at larger values of $\Bu$.

In figure \ref{subfig:ratioAp}, we plot the ratios $\frac{\SB}{\NB}$ (in dB/cycs) and $\frac{\phiaccum}{\NB}$ (which is dimensionless) which we derived using equation \ref{eq:const2chars} and are purely functions of $\Ap$. These ratios are strongly dependent on $\Ap$ and hence if we obtain reported values for either of these ratios, we can determine the value of $\Ap$. 

\subsubsection{Use of Expressions}
In this section, we have studied the dependence of filter characteristics (including characteristic ratios) on filter constants. The filter characteristics discussed here are common among the three classes of realizable filters discussed in this paper and the our findings apply to these classes as shown in a later section. Characteristics that are inherently different across the classes of filters such as the asymmetry of the power spectrum are not discussed in this paper. 

We have shown a wide range of filter characteristics accessible when fully availing the degrees of freedom of a given filter class. The full range of behavior shown in figure \ref{fig:chars} should encourage us to prioritize proper estimation of filter constants over choosing a filter class with a particular type of off-peak behavior. We also derived expressions for characteristic ratios which we later use to determine constraints on filter constant values (particularly $\Bu$) and to inform the methods by which we estimate constants given reported values of filter characteristics.

The observations we made in this section have direct implications for characteristics-based estimation of filter constants: We note that (non-ratio) characteristics must be used in combinations to estimate the filter constants. For instance, reported bandwidth-based characteristics depend on both the pole and filter exponent, and their use alone to estimate the filter constants would lead to a ill-posed inverse problem in which small changes in the $\Bu$ can be compensated by changes in the $\Ap$. In contrast, this identifiably issue does not arise when using characteristic ratios alone that depend primarily on a single filter constant.

\subsection{Methods for Estimating Filter Constants from Characteristics}

Our goal is to estimate filter constants for auditory filters mimicking human signal processing based on recent reported values of filter characteristics and characteristic ratios. This requires developing accurate methods for estimating filter constants given specifications on filter characteristics. In this section, we discuss the appropriate choices for the set of filter characteristics used for this estimation. We then present methods for accurate characteristic-based filter constant-estimation and filter design, and finally compare the proposed methods with existing characteristics-based methods.

\subsubsection{Sets of Filter Characteristics Used}

The choice of characteristics used to estimate $\bp$ is straightforward - are are typically provided with $\Bcenter$ and CF from which we may estimate the value of $\bp$. However, it is important to discuss the characteristics used to estimating the remaining filter constants, $\Ap$ and $\Bu$. 

The particular choices of characteristics and/or characteristic ratios used for estimating the filter constants depend not only on the availability of reported values, but also on the needs of a particular study or application. For instance, if we are interested in designing filters that are sharply tuned but have a limited group delay, it is natural to simultaneously specify a mixed set of two characteristics - one on a magnitude-based characteristic, and the other on a phase-based characteristic (e.g ERB and $\NB$) or an equivalent set of characteristic and characteristic-ratio (e.g. $\Qerb$ and $g$ or $\SB$ and $\frac{\SB}{\NB}$). Similarly, we would choose to use such mixed-type specifications if we are interested in dictating the frequency selectivity and synchronization  across filters in a filter bank.

Alternatively, if the application or study of interest requires fine control over the magnitude of the frequency response, then it is appropriate to simultaneously specify a set of two magnitude-based filter characteristics (e.g. $\SB$ and $\Qerb$) or an equivalent set of filter characteristic and characteristic ratio which are then used to estimate the filter constants. We suggest avoiding the combined set of $\Qerb$ and $Q_{10}$ or alternatively, one of these two characteristics along with the ratio $\alpha$. This is due to the fact that $\alpha$ varies slowly with $\Bu$  - or equivalently, $|\frac{d\Bu}{d\alpha}|$ is very large, over the region of interest (see figure \ref{subfig:ratioBu}), and hence these sets are not appropriate for reliably estimating filter constants.

 If we are designing filters to mimic auditory signal processing (as is the case here) rather than to fulfill arbitrary specifications on characteristics, we avoid the use of the $\phiaccum$ characteristic to estimate $\Bu$. This is due to the fact that the filter classes were all constructed based on the shape of tuning curves. However, these curves are only measured in the region near the peak which is what predominantly influences the results of signal processing. Consequently, while the filter classes may be used to mimic auditory signal processing of a particular species, it is inappropriate to suggest that the off-peak behavior of these filters (including $\phiaccum$) mimics that of the actual auditory system.

\subsubsection{Proposed Method}

Once we have chosen the set of characteristics to use, we proceed to the filter constant estimation method. Rather than simultaneously solving for the filter constants, we first start by estimating $\bp$, followed by $\Bu$, and finally, $\Ap$.

As mentioned above, we may estimate the value of $\bp$ from $\Bcenter$. Alternatively, if we are only concerned with a single level, we may simply set $\bp = 1$. We may then estimate $\Bu$ using reported (or computed) ratios of filter characteristics that depend solely on $\Bu$. Characteristic ratios or combinations such as $\ERBB \NB, \textrm{BW}_{3, \B} \NB, \ERBB^2 \SB, \textrm{BW}_{3, \B}^2 \SB$ are preferable to characteristic ratios such as $\frac{\textrm{BW}_{3,\B}}{\ERBB}$ or $\frac{\textrm{BW}_{3,\B}}{\textrm{BW}_{10,\B}}$ as the latter set is not particularly sensitive to variations in $\Bu$ over the parameter region of interest and hence we cannot obtain reliable estimates of $\Bu$ from these. After being provided with (or computing) $\Bu$-dependent characteristic ratios, we solve the nonlinear equation in the characteristic ratio (based on equation equation \ref{eq:const2chars}) to estimate $\Bu$. 

 We then use the estimates of $\Bu$ and $\bp$, along with the value of one of the reported characteristics (e.g. $\Qerb$), and the expression for the corresponding filter characteristic in equation \ref{eq:const2chars} to solve for $\Ap$. This is a simple linear equation that is directly solved for $\Ap$. We return to these methods and make use of them in later sections to estimate the filter constants from recent reported filter characteristics.

 We note that instead of estimating $\Bu$ followed by $\Ap$, we may do the opposite - after estimating $\bp$ we may first estimate $\Ap$ using the expressions for one of the $\Ap$-dependent characteristic ratios (e.g. $\frac{\SB}{\NB}, \frac{\phiaccum}{\NB}$) and then estimate $\Bu$. However, as previously mentioned, we avoid using designing filters based on $\phiaccum$, and note that $\SB$ are currently not reported.

\subsubsection{Comparison with Existing Methods}

The aforementioned methods of estimating the values of the $\bp$ then solving a nonlinear equation for $\Bu$ then a linear equation for $\Ap$ allows us to design filters given reported characteristics more accurately (if using $\Qerb$) than previously developed characteristic-based filter design methods [\cite{alkhairy2024characteristics}]. Previous methods involved the derivation and use of explicit expressions for each filter constants in terms of desired filter characteristics. This extra step made the constant estimation problem (and filter design methods) far more direct and did not require solving a nonlinear equation (when using $\Qerb$). However, it required making an additional approximation for the case of $\Qerb$ beyond that of the sharp-filter approximation - which allowed us to express $\log(\ERBB)$ as a linear combination of $\log(\Ap), \log(\Bu)$. This  in turn reduced the accuracy of the method compared the the methods presented in this section which require no additional approximations beyond the sharp-filter approximation used to derive equation \ref{eq:const2chars}.

As a result, if we are interested in designing auditory filters  with increased accuracy, we suggest using the methods outlined above. At the step involving solving the nonlinear equation for $\Bu$, we may provide an initial estimate for $\Bu$ based on the previous direct methods for filter constant estimation, or simply by using figure \ref{fig:chars} and \ref{subfig:ratioBu} as look-up tables to find the initial estimates.

Alternatively, if we are interested in systematically and semi-analytically studying the dependence of findings of perceptual models or technological contributions on filter characteristics, it may be more appropriate to deal with the previously derived methods which provided explicit expressions for filter constants (and hence any function of these filter constants deemed useful for these studies) directly in terms of filter characteristics.

\section{Realizable Filters}

This section establishes that the characteristic–constant relationships and methods derived in the previous section are not artifacts of a particular class, but apply broadly across commonly used gammatone-family filters. In the previous section, we worked with the transfer function $\HSharp$ which approximates the transfer function of several realizable filter classes of the Gammatone family of filters. We first reviewed the expressions for $\HSharp$ in the normalized frequency domain parameterized by filter constants. We then presented expressions for the filter characteristics of $\HSharp$ in terms of filter constants, analyzed the behavior of these filters and the sensitivity of their filter characteristics to changes filter constants in the parameter region of interest, and developed methods to design the filters (i.e. estimate the filter constants) given arbitrary specifications on filter characteristics. The filter with $\HSharp$ is itself is unrealizable but it approximates several classes of realizable filters in the peak region when $\Ap$ is small - as is the case for filters mimicking processing in humans.

In this section, we focus on the classes of realizable filters which have a transfer function that may be approximated by $\HSharp$ in the peak region. We first review these classes of filters and show that they are all parameterized by the same set of filter constants. We then demonstrate that the behavior and dependence of peak-centric characteristics describing these realizable classes of filters carry over from those obtained for the filter with $\HSharp$. Consequently, if we use peak-centric characteristics to estimate the filter constant, we only need to do so once regardless of filter class. Note that we are not interested in characteristics that differ across classes such as asymmetry or characteristics related to the magnitude at the peak.

\subsection{Classes of Realizable Filters}

The classes of realizable filters that we study here are: the Generalized Exponent Filter (GEFs/P filters), the V filter, and the classical Gammatone filter (GTFs). These are tightly related to the All Pole Gammatone Filter (APGF) and the One Zero Gammatone Filter (OZGF), as well as the Q filter and Differentiated All Pole Gammatone Filter (DAPGF) as described in Table \ref{tab:realizableClasses}. Consequently, we expect that our conclusions regarding the applicability of our conclusions and methods for $\HSharp$ not only applies to GEFs/P, V, and GTFs, but also to the other classes of filters in the table.

\begin{table*}[]
    \centering
    \begin{tabular}{|p{0.2\textwidth}|p{0.23\textwidth}|p{0.23\textwidth}|p{0.23\textwidth}|}
         \hline 
         \textbf{Summarizing Reference}
         &
         \textbf{(Repeated) pole-pair}
         & 
         \textbf{With zero I} & 
         \textbf{With  zero(s) II} 
         \\\hline
        \cite{alkhairy2024characteristics} 
        & 
        GEFs/P 
        \begin{itemize}
          \item Rational exponent
          \item Related to differential pressure in cochlear model
        \end{itemize}
        & 
        V 
        \begin{itemize}
          \item Rational exponent
          \item Related to velocity in cochlear model (if parameterized by $\Bu + 1$)
          \item Zero is equal to the negative of the real part of the pole
        \end{itemize}
        & 
        Q and generalized-Q
        \begin{itemize}
          \item Rational exponent
          \item Q has a repeated zero at zero
	   \item Generalized-Q has a repeated arbitrary zero that is far from the peak
        \end{itemize}
        \\ \hline
        \cite{katsiamis2007practical} 
        &
        APGF 
        \begin{itemize}
          \item Integer exponent
          \item Hardware (not only software) implementations
        \end{itemize}
        &
        OZGF 
        \begin{itemize}
          \item Integer exponent
          \item Arbitrary zero introduces an additional degree of freedom
          \item Hardware (not only software) implementations
        \end{itemize}
        &
        DAPGF  
        \begin{itemize}
          \item Integer exponent 
	   \item Has a single zero at zero
          \item Hardware (not only software) implementations
        \end{itemize}
        \\ \hline
        
    \end{tabular}
    \caption[Realizable filters]{Description of various classes of realizable auditory filters of the gammatone family which behave the same at the peak and for which we may estimate filter constants using the filter design methods derived for $\HSharp$ that are based on peak-centric filter characteristics}
    \label{tab:realizableClasses}
\end{table*}

\subsubsection{Generalized-Exponent (P) Filters}

The first class of realizable filters for which our conclusions from $\HSharp$ apply is the Generalized-Exponent (or P) class of filters which have a transfer function,

\begin{equation}
    \HP(s; \Ap, \bp, \Bu) \propto \bigg( (s-p)(s-\pconj) \bigg)^{-\Bu} \;.
\end{equation}

In separate work, we had shown that GEFs with rational values of the exponent, $\Bu$ may be implemented in software and used to process signals [\cite{alkhairy2025rational}]. As a result, we need not restrict ourselves to integer values of $\Bu$.

In figure \ref{fig:PandVandSharp}, we show the frequency response magnitude and phase for GEFs (blue solid lines) and the sharp-filter approximation (dotted yellow lines) for a few different sets of filter constant values of interest. The figure shows that  the bandwidth of $\HP$ is well approximated by that of $\HSharp$ - particularly when we plot the frequency response magnitude as a function of $\B - \Bcenter$.

The GEFs have an impulse response which we express in $\ttilde = 2\pi \CF t$, and which may be expressed for integer and half-integer values of $\Bu$ as follows,

\begin{equation}
    g_P \propto \expn{-\Ap \ttilde} \ttilde^{\Bu - \frac{1}{2}} J_{\Bu - \frac{1}{2}} (\bp \ttilde) u(\ttilde) \quad \textrm{for } \Bu \in \frac{\mathbb{Z}^+}{2}   \;,
\end{equation}

where $u(\ttilde)$ is the Heaviside step function and where $J_{\Bu - \frac{1}{2}} (\bp \ttilde)$ is a Bessel function of the first kind. 

We may approximate the impulse response (for $\ttilde \geq 0$) by the term,

\begin{equation}
    g_P \appropto \expn{-\Ap \ttilde} \ttilde^{\Bu - 1} \cos (\bp \ttilde - \Bu \frac{\pi}{2}) \quad \textrm{for } \Bu \in \frac{\mathbb{Z}^+}{2}   \;,
    \label{eq:gP}
\end{equation}

as discussed in \cite{alkhairy2025rational}. This form resembles GammaTone Filters (GTFs) but the oscillatory component has a phase shift that is tied to the filter exponent $\Bu$, and we have expanded the domain of $\Bu$ beyond positive integers.  We refer to this version of GTFs as P-GTFs to distinguish it from the classical GTFs with an additional degree of freedom for the phase shift. As noted by \cite{darling1991properties}, the choice of phase shift has negligible effect on the power spectrum.

We note that the transfer function of GEFs/P is essentially the same as that of APGFs but without imposing that $\Bu$ is an integer. Consequently, any conclusions we make regarding the applicability of findings and methods from $\HSharp$ to $\HP$ also applies to $H_{APGF}$. The reasons we directly work with (the notation of) GEFs/P here rather than the more commonly known APGFs is as follows:

\begin{itemize}
\item The GEFs/$P$ filters do not require that the filter exponent, $\Bu$ be an integer. This allows us to access a broader range of filter behaviors and more finely tune the filter based on desired values for filter characteristics.
\item The accuracy of previous characteristics-based filter design methods was assessed for GEFs.
\item  GEFs/$P$ filters have origins in a unified model and can be used both as auditory filters and also as part of a cochlear model to understand the mechanisms underlying a single partition box model of the cochlea (where GEF/P filters are related to the pressure difference across the Organ of Corti in the model). Consequently, having methods to design these filters given specifications on filter characteristics also allows us to study the dependence of the related cochlear model mechanisms on the observed characteristics as described in \cite{alkhairy2024cochlear}.
\end{itemize}

\subsubsection{V Filters}
Another class of filters for which the transfer function may be approximated by $\HSharp$ - and which therefore may also be designed using the same set of methods developed for characteristics-based filter design, is the $V$ class of filters which has a transfer function,

\begin{equation}
    \HV(s; \Ap, \bp, \Bu) \propto  (s+\Ap) \bigg( (s-p)(s-\pconj) \bigg)^{-\Bu} \;.
    \label{eq:HV}
\end{equation}

As seen in figure \ref{fig:PandVandSharp}, GEFs and V filters exhibit different off-peak behavior (particularly the asymmetry of the magnitude about the peak) but are quite similar in the peak region, and the transfer function for both may be approximated by $\HSharp$.

We note that V filters are similar to OZGFs but with one less degree of freedom. Consequently, any conclusions we make regarding the applicability of findings and methods from $\HSharp$ to $\HV$ also applies to $H_{OZGF}$ in aspects where the effect of the position zero is negligible. We make the choice of describing V rather than OZGFs here for the following reasons,

\begin{itemize}
\item Unlike OZGFs, it has the same degrees of freedom as $\HSharp$
\item Like GEFs, V filters are not required to have integer exponents
\item Like GEFs, V filters are associated with a unified cochlear model - $V$ is related to the upward velocity, $\mathcal{V}$, of the OoC partition at a particular location or CF as follows (noting the difference in exponent)
\end{itemize} 

\begin{equation}
    \mathcal{V}_{\CF}(s; \Ap, \bp, \Bu) \propto \HV(s; \Ap, \bp, \Bu + 1) \;.
    \label{eq:velocity}
\end{equation}

\subsubsection{Gammatone Filters}

Our conclusions from the sharp-filter approximation also extend to the classical Gammatone filters (GTFs) [\cite{holdsworth1988implementing}]. This is most commonly used auditory filter, and has an impulse response typically expressed as,

\begin{equation}
    \hGTF \propto t^{n-1} \expn{-2\pi b t} \cos (2 \pi f_c t + \phi) u(t) \;,
    \label{eq:hGTF}
\end{equation}

where $f_c$ is the tonal frequency, $n$ is referred to as the filter order (classically imposed to be a positive integer), and $b$ is often referred to as the bandwidth parameter. This may potentially be misleading as both $b$ and $n$ may be used to control the bandwidth. However, this terminology arose due to the fact that $n$ was typically held constant.

The above impulse response may be expressed in $\ttilde$ and parameterized by the same set of filter constants as those parameterizing $\HSharp, \HP, $ and $\HV$, if we use the following mapping,

\begin{equation}
    \begin{aligned}
        f_c &= \bp \CF \\
        n &= \Bu \\
        b &= \Ap \CF  \;.
    \end{aligned}
\end{equation}

We note that when comparing the impulse response of the GTF approximation of integer and half-integer GEFs (P-GTFs) - equation  \ref{eq:gP} with the impulse response of classical GTFs - equation  \ref{eq:hGTF}, and imposing $\phi = - \frac{\pi}{2} \Bu$, we find that the two are equivalent. GTFs are the most commonly used type of auditory filter making it an obvious candidate for inclusion in this paper. The GTFs have been used in a wide range of fields, including studies and applications beyond those traditionally related to auditory signals and systems - e.g. \cite{valero2012gammatone, jin2017application, matsumoto2011application}. Even in such applications, the values of the filter constants used are typically those historical ones used for human auditory filterbanks.

We next investigate the applicability of our findings and the suitability of the characteristics-based filter design methods developed based on the sharp-filter approximation for the cases of GEFs, V, and GTFs. The transfer functions of the aforementioned classes of filters are particularly similar in the region of the peak which is where most of the power is. This is particularly true for small $\Ap$ - i.e. when the sharp-filter approximation holds. Consequently, we expect that our conclusions from $\HSharp$ carry over to these classes of filters. We also expect our conclusions to directly carry over to APGFs which is the same as GEFs. The conclusions also apply to filter classes that have different slopes and symmetries of the magnitude such as OZGFs, DAPGFs, and Q filters but only for specific ranges values of filter constants such that there is negligible effect on the peak region (e.g. the zero is far away enough from the peak).

\subsection{Behavior and Filter Design for Realizable Filters}

In previous sections, we discussed the range of possible behaviors and methods for estimating the filter constants for $\HSharp$. However, our goal is the study of behavior and estimating filter constants for the realizable classes of auditory filters. Therefore, it is necessary to demonstrate that our findings and methods for $\HSharp$ also apply for the realizable classes of filters. We focus on the errors in the filter characteristics for each of the realizable filters as a function of the filter constants. There is no need to also discuss the errors in the filter design method as they apply to realizable filters as there are no additional approximations between these expressions needed to estimate the filter constants from the filter characteristics.

\subsubsection{Applicability for GEFs/P}

We first discuss the validity of the findings (equation \ref{eq:const2chars}) as they apply to the case of GEFs. Over a range of filter constants where $\Ap$ is small - including the region we expect to be most appropriate for filterbanks mimicking auditory signal processing in humans, $\HSharp$ is a good approximation of $\HP$ as seen in figure \ref{fig:PandVandSharp}.

The approximation of the phase of $\HP$ by $\phase{\HSharp}$ is more accurate than the approximation of the $\level{\HP}$ by $\level{\HSharp}$ as may also be seen from figure  \ref{fig:PandVandSharp}. Consequently, our closed-form expressions (equation \ref{eq:const2chars}) for phase-based characteristics such as $\NB$ and $\phiaccum$ are generally more accurate than our expressions for magnitude-based filter characteristics such as $\Qerb$ and $\Qn$ - though the expressions for bandwidths are more accurate than the quality factors which are affected due to the errors in $\Bcenter$. Expressions for measures of sharpness of tuning that are more peak-centric are more accurate than those that relate to bandwidths for larger $n$.

Upon numerically computing the relevant characteristic ratios from the frequency response of $\HP$, we find that the ratios are indeed approximately univariate in either $\Ap$ or $\Bu$ for the range of $\Ap$ and $\Bu$ of primary interest. We also find that, quantitatively, the numerically computed ratios are well-approximated by those in figure \ref{fig:ratios} and our closed-form expressions in equation \ref{eq:const2chars}.

\begin{figure*}
    \centering
    \includegraphics[width=\linewidth]{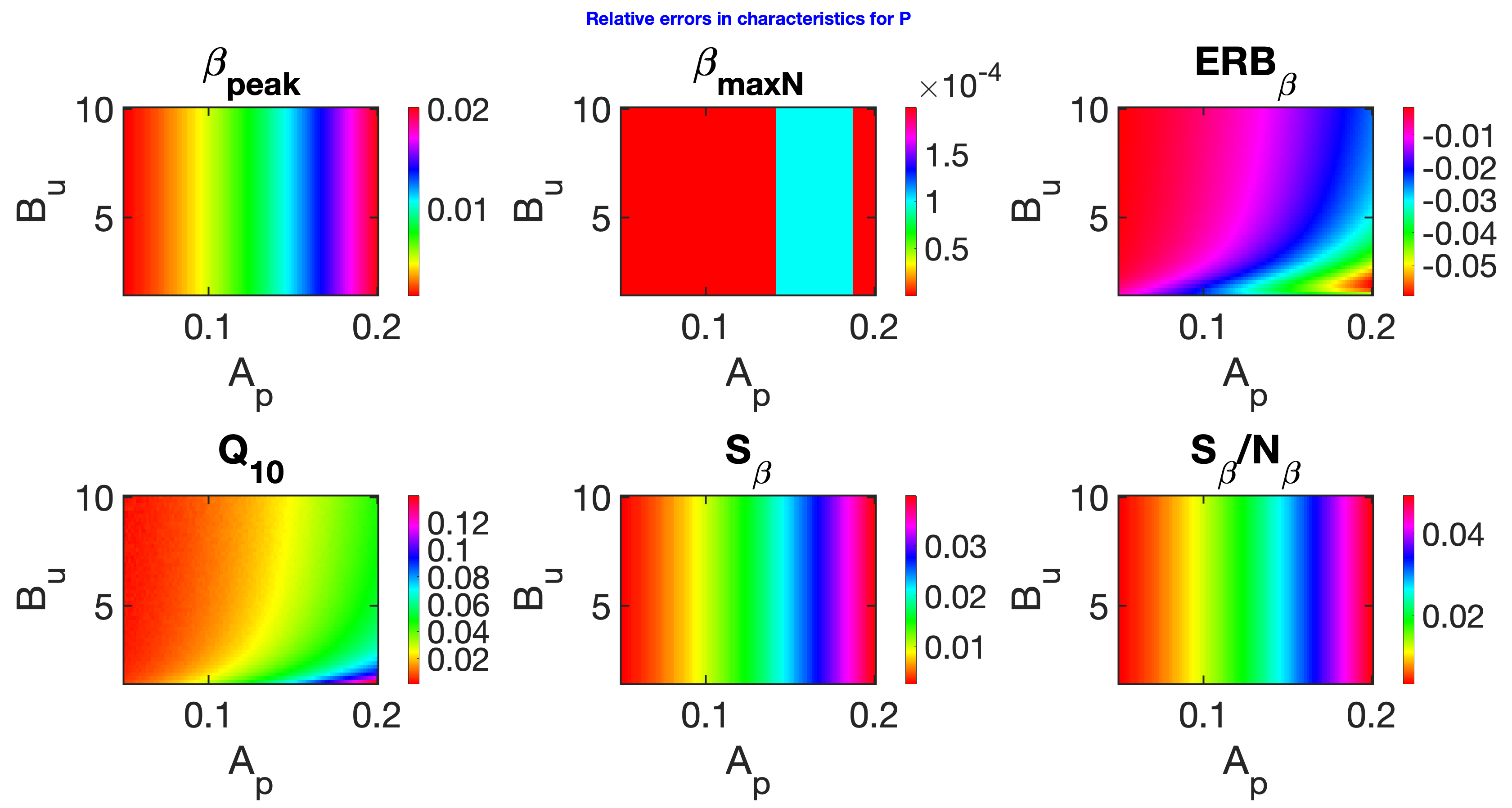}
    \caption[Error in estimated filter constants for GEFs]{We show the relative errors (as a function of filter constants $\Ap$ and $\Bu$ with a fixed $\bp = 1$) of the values filter characteristics obtained using our closed-form expressions (that are based on the sharp-filter approximation) compared to those computed numerically from $\HP$. The relative errors are small as may be inferred from the colorbars, indicating the accuracy - for the case of GEFs, of our expressions for filter characteristics in terms of filter constants, and consequently, the accuracy of filter design methods that depend on it.
    The errors typically increase with $\Ap$ - i.e. as the sharp-filter approximation starts to break down. }
    \label{fig:errCharsP}
\end{figure*}

We demonstrate the accuracy of expressions for characteristics in terms of constants for the case of GEFs/P in figure \ref{fig:errCharsP} which shows the relative error in a set of filter characteristics as a function of the filter constants $\Ap$ and $\Bu$ (while fixing $\bp = 1$). We define the relative error in a particular filter characteristic, $\gamma$, at some value of $(\Ap, \Bu, \bp = 1)$ as the deviation from unity of the ratio of $\gamma_{\textrm{num-P}}$ to $\gamma_{\textrm{analytic}}$ where $\gamma_{\textrm{num-P}}$ is the the value of $\gamma$ obtained numerically from the frequency response of GEFs that were designed with the specified values of filter constants, and $\gamma_{\textrm{analytic}}$ is the value of $\gamma$ obtained based on the sharp-filter expressions (equation \ref{eq:const2chars}) using the same specified values of filter constants. We do not take the absolute value in order to determine when we are consistently over- or under-estimating certain filter characteristics. For instance, the error in $\Qerb$ is defined to be $\varepsilon_{\Qerb,P} \triangleq 1 - \frac{Q_{\textrm{erb}, num-P}}{Q_{\textrm{erb, analytic}}}$. The resultant relative errors shown in figure \ref{fig:errCharsP} demonstrate the appropriateness of using the expressions of equation \ref{eq:const2chars} for determining the behavior of GEFs and estimating their filter constants from specifications on filter characteristics.

\subsubsection{Applicability for V and P-GTFs}

\begin{figure*}
    \centering
    \includegraphics[width=\linewidth]{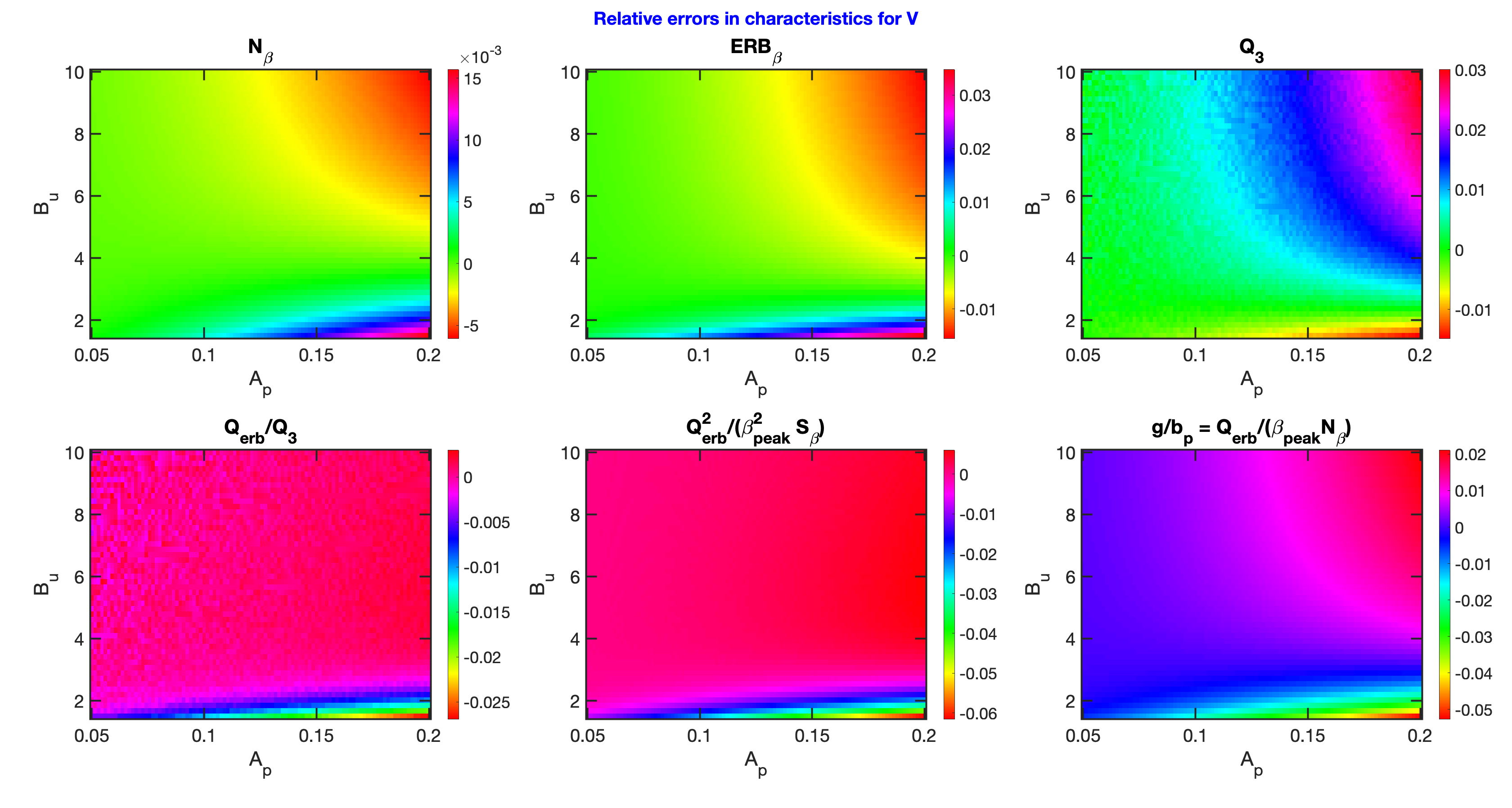}
    \caption[V errors]{Relative errors in filter characteristics are small for the case of V filters as seen from the colorbars, indicating the accuracy of our expressions and methods for this class of filters.}
    \label{fig:errCharsV}
\end{figure*}

\begin{figure*}
    \centering
    \includegraphics[width=\linewidth]{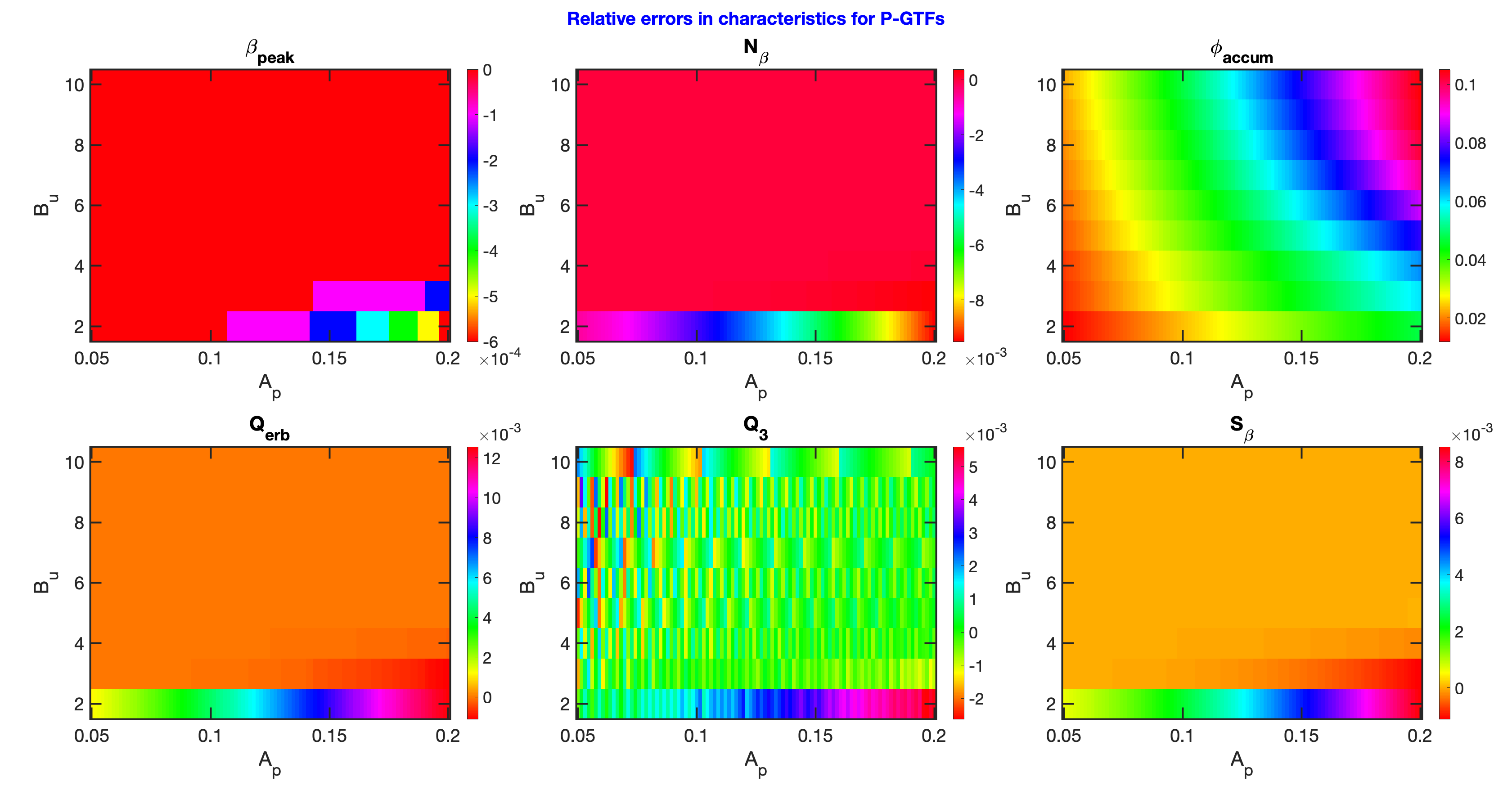}
    \caption[GTF errors]{Relative errors in filter characteristics are small for the case of P-GTFs (GTFs with a phase shift of $-\Bu \frac{\pi}{2}$ which is derived based on the approximation of the impulse response of GEFs except at the lowest times) as seen from the colorbars, indicating the accuracy of our expressions and methods for this class of filters.}
    \label{fig:errCharsGTF}
\end{figure*}

We also show the relative errors in two of the other classes of realizable filters, $V$ and $P-GTFs$ (GTFs with $\phi = - \Bu \frac{\pi}{2}$) \footnote{As previously mentioned, \cite{darling1991properties} has shown that the value of the phase shift has negligible effect on the frequency response magnitude }, in figures \ref{fig:errCharsV} and \ref{fig:errCharsGTF} respectively.

In figure \ref{fig:errCharsV}, we show that the relative error in various filter characteristics for the case of $\HV$ is small - particularly for smaller values of $\Ap$ as required for the validity of the sharp-filter approximation assumption. The transfer function for $V$ has a zero at $-\Ap$ (with an $\Ap$ that is particularly small for humans). Therefore, when compared with $\HP$, the symmetry of the magnitude of $\HV$ is in fact closer to that of $\HPsharp$. Consequently, while we had derived $\HSharp$ as an approximation of $\HP$, the derived expressions for filter characteristics as a function of filter constants are in fact generally more accurate for $\HV$ than they are for $\HP$.

In figure \ref{fig:errCharsGTF} we show the relative errors in certain filter characteristics for the case of P-GTFs . Previous work had derived $\HSharp$ as an approximate transfer function for $\hGTF$  when the sharp-filter approximation holds [\cite{holdsworth1988implementing, darling1991properties}]. Indeed, we find that the errors in figure \ref{fig:errCharsGTF} are small over the parameter region of interest.

\section{Constraints and Estimates for Filter Exponent}

In previous sections, we studied the dependence of filter characteristics on filter constants which is integral to determining which filter characteristics may reliably be used to estimate filter constants. We also presented methods to estimate filter constants from sets of filter characteristics. 

In this section, we take one step further towards estimating filter constants by deriving constraints on, and estimates for, the filter exponent $\Bu$ based on a variety of qualitative and quantitative observations. We summarize these constraints and estimates in Table \ref{tab:BuVals}. As previously mentioned, we the methods we follow to estimate the filter constants involve first estimating $\Bu$ (carried out in this section) followed by $\Ap$ (done in the next section). The various constraints discussed below serve different roles: some provide bounds, others provide point estimates, and others simply suggest approximate CF-invariance of the filter exponent.

\begin{table*}[htbp]
    \centering
    \begin{tabular}{lp{0.6\linewidth}p{0.22\linewidth}}
    \hline
    
    \textbf{\#} & \textbf{Data} & \textbf{$\Bu$ Condition/Value} \\  \hline

    N/A & condition on envelope of impulse response for half-integer domain & $\frac{3}{2}\leq \Bu$ \\

    N/A & range of $\alpha$ (physiological, across lab species)  & $3.86\leq \Bu$ \\
    
    $B_{u,0}$ & historical (designed GTFs based on simultaneous masking psychophysical experiments in humans) & $\Bu = 4$ (more generally, $\Bu \in \{3,4,5\}$)\\

    $B_{u,1}$ & $g=g_1$ (chinchilla physiological data - specifically ANF WKs) & $\Bu \approx 7.2$ \\

    $B_{u,2}$ for $\CF > \CF_{a|b}$& $g = g_2$ (from $r$ found to be species-invariant with that of humans from $\Qerb$ from forward masking psychoacoustic experiments and human $\Nsfoae$ using the paradigm of \cite{shera2010otoacoustic}; and from the (conservative) range of $\eta$ across species) & $ 2 \leq \Bu $ \\
    \hline
    \end{tabular}
    \caption[Conditions and estimates of filter exponent]{The table summarizes conditions on, and estimates of, the filter exponent based on historical extensions and based on reported values of characteristic ratios. Conditions regarding the approximate independence of $\Bu$ on CF are not mentioned in the table but are detailed in the main text.}
    \label{tab:BuVals}
\end{table*}


\subsection{Historical Estimates}

The historically chosen value for $\Bu$ based on fitting the transfer function to the frequency response magnitude from simultaneous masking psychoacoustic experiments in humans is $B_{u,0} = 4$. The range of $\Bu$ deemed to be desirable historically based on the same simultaneous masking experiments is $3 \leq \Bu \leq 5$)\footnote{Strictly, speaking it was the set of integers, $\Bu \in \{3,4,5\}$ that were historically studied and found to be appropriate based on fitting to tuning curves from simultaneous masking experiment - often via Roex filter fits}. These are based on simultaneous masking experiments which were performed decades ago, and involve fitting to tuning curves. 

If we are provided with a tuning curve - regardless of how it was measured, and we are only concerned with fitting to that tuning curve, then the most appropriate method of estimating filter constants is to fit to the tuning curve using classical filter design methods that take a frequency response magnitude over a length of frequencies and returns estimated filter constants. This is appropriate provided that individual data from the psychoacoustic tuning curves are reliable and not when we need to average over trials, subjects, or experiments or when recorded curves deviate from the shape of the filter classes. In these cases using characteristics-based methods are preferable to fitting.

More importantly, fitting to the tuning curve leaves us with a single set of filter constants specific to a certain case, and without intuition of how the constants relate to the filter behavior. Furthermore, reported data is often in the form of filter characteristics. In what follows we use focus on constraints and estimates of $\Bu$ that we can obtain based on filter characteristics. We focus on those observations relevant for designing auditory filters that mimic signal processing in the healthy human, but also arrive at conditions that apply regardless of species. Additionally, our analysis below more generally provide an example of how to use the characteristics-based design methods -  given arbitrary specifications on filter characteristics or given reported specifications, present constraints based on observations, and identify which filter characteristics (including ratios) are most informative.


\subsection{Conditions on the Variation of Filter Constants with CF}

The first set of conditions we arrive at for $\Bu$ are with regards to its variation with respect to CF. 

\subsubsection{Trend in Quality Factor}

One such condition occurs due to the qualitative trend in reported quality factor. Our transfer functions are in normalized frequency, $\B$, and we define our filter constants accordingly. We note that fixed values of $\Ap, \bp, \Bu$ across filters in a filterbank result in the constant-Q filters (in other words, the bandwidth defined in $f$ increases proportionally with $\CF$). However, experimental observations suggest that the quality factor is larger for larger CF [\cite{robles2001mechanics, glasberg1990derivation}]. Based on equation \ref{eq:const2chars} and figure \ref{fig:chars}, this means that the value of $\Ap$ should be smaller for larger-CF filters and/or the value of $\Bu$ should be larger for larger-CF filters. 

\subsubsection{Condition from Traveling Wave Model}

The second set of observations are also regarding how we expect the filter constants to vary with CF. Let us return to the origin of GEFs and V which provided a unified framework for a filterbank and a single-partition box model of the cochlea. The model tied together the traveling wave and transfer function perspectives. As  the energy inputted into the cochlea from the stapes is expected to predominantly travel in the forward direction (i.e. from the base to the apex). This translates to a requirement that the model constants parameterizing the wave number vary slowly with $x$ \footnote{We note that this condition is most easily derived for another class of models - 1D two-way wave equations (such as those describing transmission line models) and is referred to as the WKB condition. Our cochlear model is a short-wave model collapsed from 2D into a one-way 1D wave equation which requires a separate derivation for the mathematical condition which is therefore beyond the scope of this filter-design paper.} Consequently, we expect that the filter constants which parametrize the transfer functions - and which are the same as the model constants in this case, must vary slowly with CF.

\subsubsection{Condition on Polarity}

A tighter condition on the variation of $\Bu$ with respect to CF comes from an observation regarding the polarity and zero crossing of auditory nerve fiber (ANF) click responses which were found to be the same across the length of the cochlea [\cite{Guinan}]. It is not clear why this occurs from a cochlear mechanism perspective as the value of $\Bu$ in the cochlear model is implicitly a function of material and geometric properties which \textit{do} vary with $x$. Nevertheless, this observation provides a polarity condition on the oscillatory component of the impulse response of most filter classes. This is true for the sharp-filter approximation, GEFs, $V$ filters, and P-GTFs but not for classical GTFs where the phase shift of the tonal component is not tied to $\Bu$. Coupled with the above condition that the filter constants must vary slowly with CF translates into a condition that $\Bu$ is approximately constant with respect to CF. The approximate CF-independence of $\Bu$ also translates into approximately constant values for $\Bu$-characteristic ratios - such as those in equation \ref{eq:charratios} and figure \ref{subfig:ratioBu}, and further discussed in later sections. Because this condition is dependent on the filter class - or more specifically, on the filter exponent being tied to the phase shift of the tonal component of the impulse response as applied to the actual ANFs rather than filters we are attempting to design, we keep in mind that this condition may not necessarily hold. However, we later separately arrive at a condition on $\Bu$ from reported $g$ that also suggests that $\Bu$ is CF-invariant, and hence we proceed with this conclusion.

\subsubsection{Observations of Phase Accumulation}

The constant nature of $\Bu$ means that the phase accumulation is constant across CF. However, we currently do not have measurements of frequency response phase much away from the peak. Furthermore, the filter classes were constructed to fit responses primarily near the peak frequency and were not tested beyond it where the magnitude is negligible and data is not reliable. As a result, we do not give much credence to observations regarding phase accumulation with the exception of using the partial phase accumulated over the peak region which provides a lower bound for phase accumulation and hence for $\frac{\Bu}{2}$.

\subsubsection{Existing Reports of Variation}

The above conclusion that $\Bu$ is CF-invariant appears to contradict previous estimates of $\Bu$ for chinchilla reported in \cite{alkhairy2019analytic}. Those estimates were found by fitting $\mathcal{V}_{\CF}$ of equation \ref{eq:velocity} (which is related to GEFs and V filters but with an exponent of $\Bu + 1$) to Wiener Kernel (WK) data from chinchilla ANFs. Specifically, \cite{alkhairy2019analytic} showed that the trend lines for $\Bu + 1$ ranged from about $4$ for the lowest CF to about $6$ for the highest CFs. However, we make the following notes about the previous findings: (1) there was a very large spread in estimated values of $\Bu$ from ANF WKs - from about 1.75 to 7.75 - this included a large spread in $\Bu$ estimated from WKs collected at the same CF from different chinchilla which implies that the trend lines alone provided insufficient information, (2) the ANF WK data included those that were very noisy at the peak, (3) while the sharp-filter approximation holds for the chinchilla base, it is not as reliable below a CF of $5$ kHz in chinchilla, and (4) a lower bound on $\Ap$ was artificially imposed which affected the estimates for $\Bu$. As a result, we build on the finding that $\Bu$ is approximately CF-invariant.


\subsection{Constraint Due to Shape of Impulse Response Envelope}

The shape of the envelope of the impulse response provides a lower bound on the value of $\Bu$. The impulse or click responses in alive animals across species and CFs may roughly be described by a tone modulated by an envelope that grows at the earlier times then decays - rather than having a pure decay (we are not concerned with frequency glides or phenomenon of waxing and waning here). Based on the impulse response approximation for GEFs with integer and half-integer $\Bu$ (equation \ref{eq:gP}), this translates to a constraint, $\Bu \geq \frac{3}{2}$ as $\Bu =1$ leads to an impulse response with a pure decay. This also applies to the other filter classes as their impulse responses can also be approximated by that of equation \ref{eq:gP}.


\subsection{Constraints from Characteristic Ratios}

The final type of constraints come from reported values and ranges of ratios of filter characteristics. The expressions for some of the $\Bu$-dependent characteristic-ratios (specifically $\alpha$ and $g$) were provided in equation \ref{eq:charratios}.

\subsubsection{Ratio of Quality Factors}

\begin{figure}
    \centering
    \includegraphics[width = \linewidth]{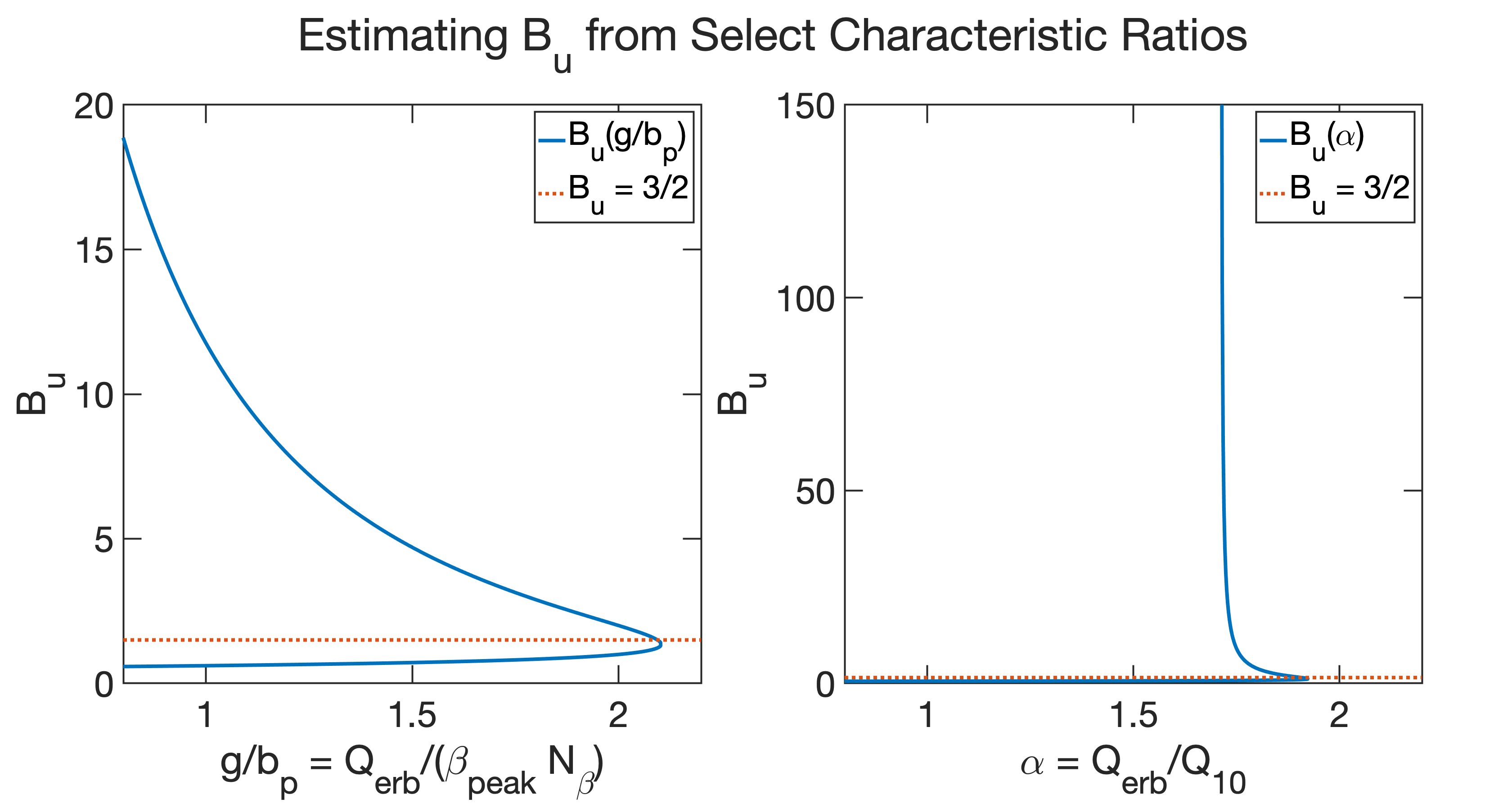}
    \caption[Bu vs ratios]{$B_u$ is plotted (blue solid line) vs two characteristic ratios ($\frac{g}{\bp}$ and $\alpha$). A line at $\Bu = \frac{3}{2}$ is included (red dashed line) to indicate a lower limit on acceptable $\Bu$. If we include this constraint, we may express $\Bu$ as a function of the filter characteristic ratios. We note that especially for $\alpha < 1.75$, there is little value in this reported characteristic ratio as far as estimating $\Bu$. This is due to the fact that in this range, $|\frac{d \Bu}{d \alpha}|$ is relatively  large and hence small differences in reported values of $\alpha$ can result in wildly different values for $\Bu$ and hence we cannot arrive at credible estimates for $\Bu$ from $\alpha$ in this range. The same holds true for smaller values of $\frac{g}{\bp}$ (occurring at larger values of $\Bu$) though to a lesser extent.}
    \label{fig:BuFromRatios}
\end{figure}

The first observation is regarding the ratio $\alpha \triangleq \frac{\Qerb}{Q_{10}}$ of equation \ref{eq:charratios}. \cite{shera2003stimulus} found that $1.7 \leq \alpha \leq 1.8$ across species and CFs. Using $1.7 \leq \alpha \leq 1.8$ and based on equation \ref{eq:charratios}, we arrive at a condition on the range of possible values for the filter exponent: $3.86 \leq \Bu$. Figure  \ref{fig:BuFromRatios} shows the dependence of $\Bu$ on $\alpha$ and illustrates that $\alpha$ in this range should only be used to provide a lower bound and cannot be used to reliably estimate a specific value of $\Bu$.  

Note that another possibility for the bounds on $\Bu$ may be inferred from $1.7 \leq \alpha \leq 1.8$. This other possibility is that $0.73 \leq \Bu \leq 0.8$. However, this is inconsistent with the earlier minimum condition we imposed on $\Bu$ to guarantees a growth then decay in the envelope of the impulse response of GEFs, and also contradicts other observations detailed later in this paper.

\subsubsection{Ratio of Quality Factor to Peak Group Delay}

Another characteristic ratio which we have reported data for and which provides a constraint on $\Bu$ is the ratio $g \triangleq \frac{\Qerb}{\NB}$ which was previously defined in equation \ref{eq:charratios}. In the following discussion, we impose $\bp = 1$ so that $g$ is purely a function of $\Bu$. Reported values of $g$ may provide constraints on (or estimates of) $\Bu$. In figure \ref{fig:BuFromRatios}, we show how $\Bu$ estimates vary with $g$. In what follows, we discuss two sets of observations related to $g$ - which we refer to $g_1$ and $g_2$, that lead to constraints on $\Bu$.

\subsubsection{First Estimate}

One reported value for $g$ comes from \cite{shera2010otoacoustic} who computed $g$ as the ratio of $\Qerb$ and $\NB$ that were both computed from frequency response curves of ANF WKs at low SPLs in chinchilla. We refer to this estimate of $g$ as $g_1$ which was reported to be $g_1 \approx 1.25 (\pm 0.02$) along the length of the cochlea. Solving equation \ref{eq:charratios} for $\Bu$ (and taking the solution that is greater than unity), we arrive at an approximately CF-invariant value of $B_{u,1} \approx 7.2$ from $g_1$. 

We note that we had previously computed an estimate of $\Bu$ from $g_1$ in \cite{alkhairy2024cochlear} using a more long-winded approach from a ratio $\nu \triangleq \frac{Q_{10}}{\NB} \equiv \frac{g_1}{\alpha}$. This was done because - at the time, we had derived expressions for $\NB$ and $Q_{10}$ but not for $\Qerb$. We had additionally reported $\Bu$ as the closest integer and arrived at $\Bu = 7$. This is consistent with our current estimate for $\Bu$ from $g_1$ of $B_{u,1} \approx 7.2$. We later use this estimate of $B_{u,1}$ along with two sets of reported  $\Qerb$ to estimate the filter constant, $\Ap$.

\subsubsection{Second Estimate}

We may also obtain estimates for $\Bu$ using a second estimate of $g$ which we refer to as $g_2$. We compute this estimate for $g$ as $g = r \eta$ from the tuning ratio, $r$, and a ratio, $\eta$. 

The tuning ratio, $r$, was defined by \cite{shera2010otoacoustic} as,

\begin{equation}
    r \bigg(\frac{\CF}{\CF_{a|b}} \bigg) \triangleq \frac{\Qerb}{\Nsfoae}\;,
\end{equation}

where $\CF_{a|b}$ is the CF at the apical-basal transition and $\Nsfoae$ is the normalized group delay of SFOAEs at a frequency corresponding to the CF of the filter of interest.

We define the ratio $\eta$ as,
\begin{equation}
    \eta \triangleq \frac{\Nsfoae}{\NB} \;.
\end{equation}

In what follows, we discuss reported values of $r$ and $\eta$, then compute $g$ as $g = r \eta$ to then provide an alternate condition for $\Bu$ based on these reported values. 

The tuning ratio $r$ has previously been reported across various species - including humans. The first reported value for $r$ which we refer to as $r_2$ was computed using $\Nsfoae = N_{\textrm{sfoae,shera}}$ with SFOAEs gathered and processed as described in \cite{shera2010otoacoustic}. The $\Qerb$ used to compute $r_2$ for most species was based on physiological experiments. In the case of humans, $\Qerb$ was computed from noninvasive experiments. More specifically, it was based on a forward masking paradigm described in \cite{oxenham2003estimates} to provide $\Qerb = Q_{\textrm{forw}}$. This study reported that the $r_2 \bigg(\frac{\CF}{\CF_{a|b}} \bigg)$ computed in this manner is approximately species-invariant.

From figure 9 of \cite{shera2010otoacoustic} we obtain an estimate $ 0.8 \leq r_2 \bigg(\frac{\CF}{\CF_{a|b}} \bigg) \leq  1$ for $\frac{\CF}{\CF_{a|b}} > 1$. We note that other studies provide (or may be used to compute) values for $r$ that appear rather inconsistent with the above estimates  - e.g. \cite{bentsen2011human}, \cite{lineton2009comparing}, and \cite{schairer2006use}. We expect that these differing estimates of $r$ and any differences between humans and other species are largely due to differences in the paradigms for obtaining SFOAEs and psychoacoustic tuning curves. This leads to an important cautionary note when estimating the set of filter constants using values of $r$ together with another filter characteristic  (e.g. $\Nsfoae$ or $\Qerb$). In particular, we must we ensure that the reported values used are consistent in terms of definitions and methodologies.

As for $\eta$, we note that while a tighter condition on $\eta$ exists based on figure 7 of \cite{shera2010otoacoustic}, we use a relatively weak condition that $1 \leq \eta_2 \leq 2$ \footnote{We do this in consideration of unresolved questions regarding the deviation of this ratio from the value of 2 originally expected based on coherent reflection theory, and because these values found for chinchilla seem to differ from those previous reported for cats and guinea pigs \cite{shera2003stimulus}}. From $r_2$ and $\eta_2$, we obtain a second estimate for $g$. Specifically, $0.8 \leq g_2 \leq 2$. Computing the range of $\Bu$ from equation \ref{eq:charratios} (and taking the values that are greater than unity), we arrive at an approximate condition $2 \leq B_{u,2} \leq 18.85$ for $\CF > \CF_{a|b}$. The upper bound is unreliable as $\Bu$ is very sensitive to small changes in the value of $g$ in this region of parameters (in contrast to the region prescribed by $g_1$) and hence not included in Table \ref{tab:BuVals} or used towards estimating the filter constant $\Ap$.

\subsubsection{Historical Value of Exponent}

For reference, here we discuss the consistency of the historically chosen values of $\Bu$ with reported (and computed) values or ranges of $\alpha$ and $g$. Consistency does not imply that the historical values of $\Bu$ provide the `correct’ solution, but rather that they are one of many possible consistent solutions.

Recall the the reported range of $\alpha$ is $1.7 \leq \alpha \leq 1.8$ which lead to a condition, $3.86 \leq \Bu$. The historically chosen value for $\Bu$ in humans ($B_{u,0} = 4$) corresponds to $\alpha = 1.8$. The range of $\Bu$ deemed to be desirable historically based on the same simultaneous masking experiments ($3 \leq \Bu \leq 5$) corresponds to $1.78 \leq \alpha \leq 1.82$.

We note that $g_1$ (based on physiological observations from chinchilla and expected to hold in humans) results in a value for $\Bu =  B_{u, 1} = 7.2$ that is greater than the historically imposed value of $\Bu = B_{u, 0} = 4$ based on simultaneous masking psychoacoustic experiments for humans. On the other hand $B_{u, 0}$ is within the wide range for $B_{u, 2}$ prescribed by the species-invariant value of $g_2$ for $\CF > \textrm{CF}_{a|b}$. The $g$ corresponding to the historical $B_{u, 0} = 4$ has a value $g = g_0 = 1.6$, and $\Bu$ in the range historically found to be desirable, $3 \leq \Bu \leq 5$ corresponds to $1.46 \leq g \leq 1.78$.

\section{Estimates for Filter Pole}

In the previous section, we used observations to estimate the value (or range of values) of the filter exponent $\Bu$. Here, we provide estimates for the last filter constant, $\Ap$, to be used for human auditory filterbanks given observations on some sets of filter characteristics (including characteristic-ratios). We then discuss limitations of our methods and estimates. 

In choosing the set of observations used to estimate $\Ap$ (and before that, $\Bu$), we bear in mind our application of interest, as well as the availability of reliable datasets, and the sensitivity of the estimates to the observed characteristics.

Following the methods described earlier, we first estimated $\bp$ then $\Bu$ and finally estimate $\Ap$. We set the peak frequency to occur at CF and hence set $\bp = 1$, and consider two estimates of $\Bu$: $B_{u,0} = 4$ which is the historically used value based on fits to simultaneous masking curves, and $B_{u,1} = 7.2$ which was obtained from $g_1 = 1.25$  \footnote{Which we assume to be species invariant and extrapolate to humans based on the arguments in \cite{shera2003stimulus, shera2010otoacoustic}}.  

\begin{figure}[htbp]
\centering

\begin{tikzpicture}[
    scale=0.8,
    transform shape,
    node distance=2.3cm,
    every node/.style={inner sep=0pt},
    arrow/.style={->, thick}
]

\node (A) {$\bp = 1$};
\node (B1) [right=of A, yshift=1cm, xshift = 2cm] {$B_{u,0} = 4, A_{p, 0} = 0.0252\frac{4.37 \CF + 1}{\CF}$};
\node (B2) [right=of A, yshift=-1cm] {$B_{u,1} = 7.2$};

\node (C21) [right=of B2, yshift=0.8cm] {$A_{p, 1-1} =0.0354 \frac{4.37 \CF + 1}{\CF}$};
\node (C22) [right=of B2, yshift=-0.8cm] {$A_{p, 1-2} = 0.1303 \CF^{-0.27}$};

\draw[arrow] (A) -- node[above] {sim, $Q_{\textrm{sim}}$} (B1);
\draw[arrow] (A) -- node[above] {$g_1$} (B2);


\draw[arrow] (B2) -- node[above] {$Q_{\textrm{sim}}$} (C21);
\draw[arrow] (B2) -- node[above] {$Q_{\textrm{forw}}$} (C22);

\end{tikzpicture}

\caption{Process to arrive at estimated constants for human auditory filters as well as estimated values for the filter constants. CF is in kHz and the filter constants are dimensionless.}
\label{fig:tree}
\end{figure}
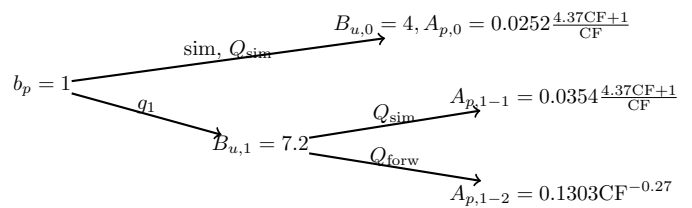

\begin{figure}[htbp]
    \centering
    \includegraphics[width=\linewidth]{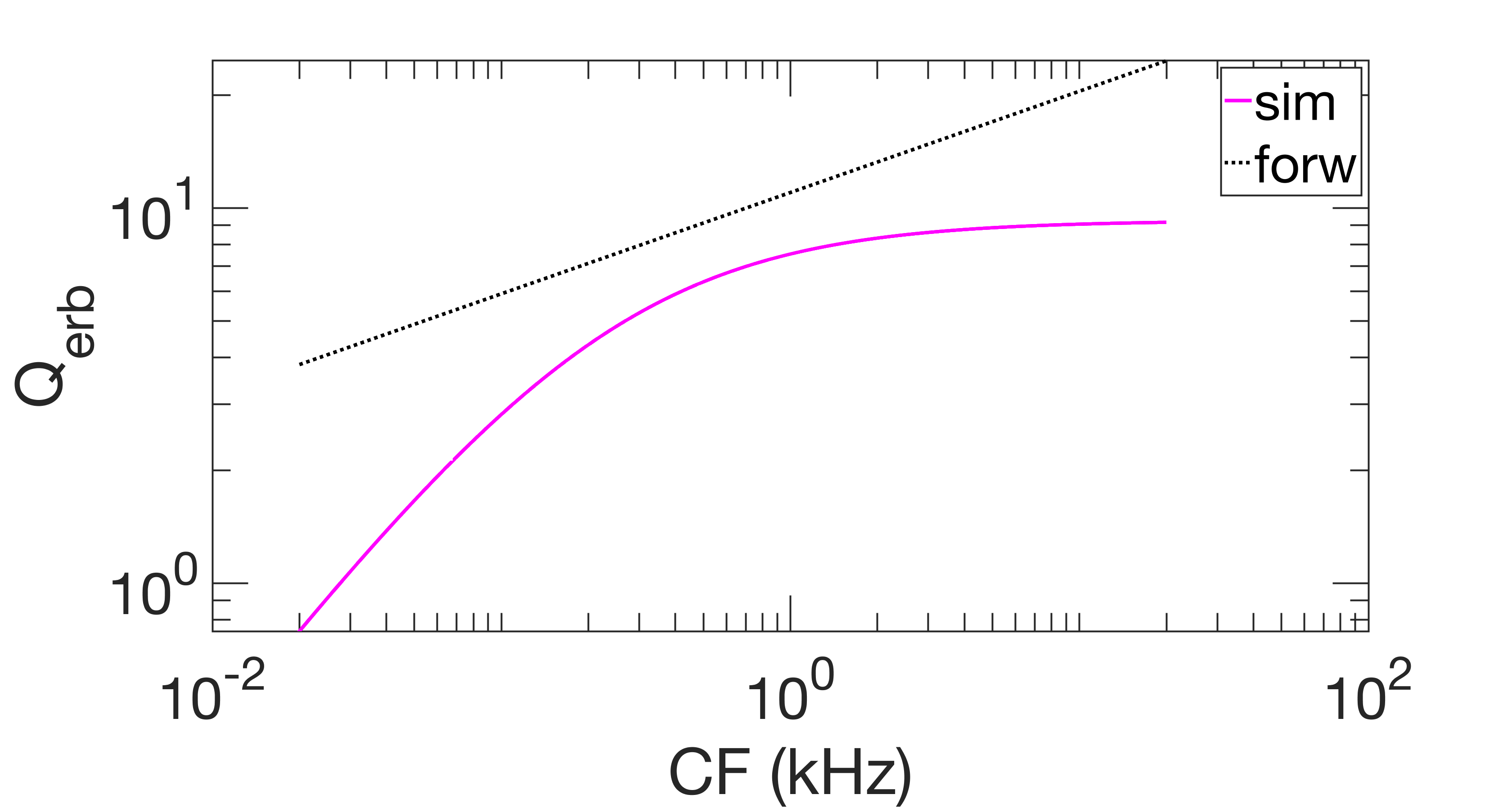}
    \caption[Qerb used]{Reported values of $\Qerb$ as a function of CF. We use the values of $\Qerb$ to estimate the filter constant $\Ap$ given estimates of $\Bu$ and $\bp$. The $\Qerb$ formulae used are based on simultaneous (magenta solid line) and forward masking paradigms (black dotted line). Both were extrapolated to cover the range of CF for humans.}
    \label{fig:reportedQerb}
\end{figure}

\begin{figure}
    \centering
    \includegraphics[width=\linewidth]{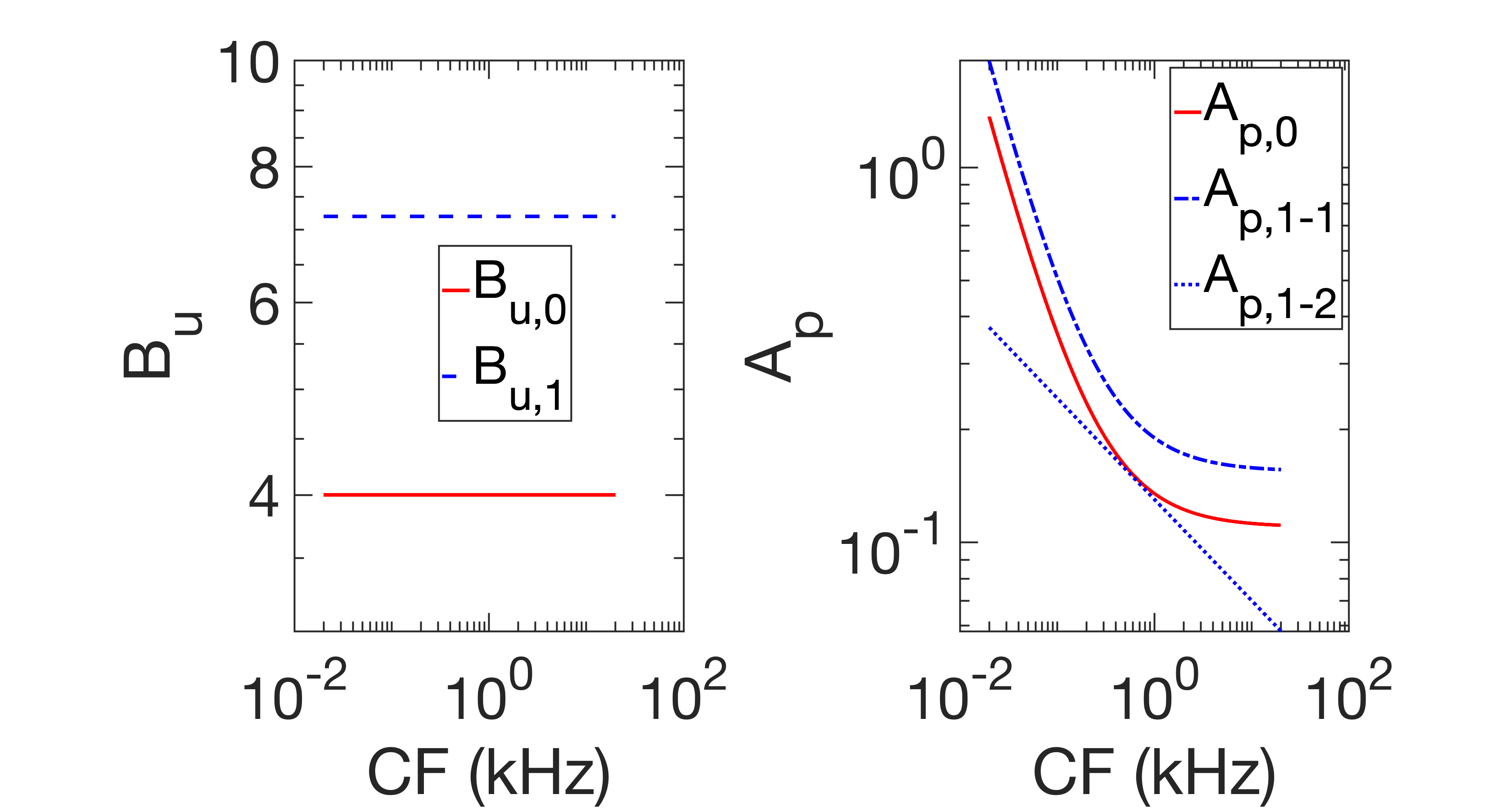}
    \caption[Estimates of filter constants using various sets of information]{Estimated values for filter constants for filterbanks mimicking processing in humans. The estimates are based on a few reported sets of values for filter characteristics including characteristic ratios as detailed in the main text. For all these estimates, $\bp = 1$. In the left panel, estimate $B_{u,0}$ (red solid) is the historical choice, and the estimate $B_{u,1}$ (blue dashed) is obtained by extrapolating $g=g_1$ to humans. In the right panel, the estimate $A_{p,0}$ (red solid) is based on the historical $Q_\textrm{sim}$ and $B_{u,0}$. The estimate $A_{p,1-1}$ (blue dash-dots) is computed from $B_{u,1}$ and $Q_\textrm{sim}$, and the estimate $A_{p,1-2}$ (blue dots) is computed from $B_{u,1}$ and $Q_\textrm{forw}$.}    \label{fig:constEstVsCF}
\end{figure}

We then take each of these values of $\Bu$ along with reported values of $\Qerb$ (shown in figure \ref{fig:reportedQerb} \footnote{Note that 
the reported values of $\Qerb$ are over a particular CF, but we have extended this to a wider range of CFs for figure \ref{fig:constEstVsCF}}) to derive expressions for $\Ap$ as a function of CF as schematized in figure \ref{fig:tree}. In figure \ref{fig:constEstVsCF}, we show the various resultant estimates for the filter constants for human auditory filters as a function of CF. In estimating $\Ap$, we ensure that the quality factors used are consistent with the $\Bu$. For example, to estimate $\Ap$, we cannot combine quality factors obtained from forward masking experiments with estimated values for $\Bu$ obtained from fitting filters to tuning curves from simultaneous masking experiments. We note that unlike the case of $B_{u,0}$ which is tightly associated with $Q_{\textrm{sim}}$, we do not know what the appropriate choice of $\Qerb$ is for $B_{u,1}$ based on $g_1$. For this reason, we compute two estimates for $A_{p,1}$ using $B_{u,1}$ and two possible choices of $\Qerb$.

\subsection{From Tuning Curves of Simultaneous Masking Experiments}

For reference, we start by providing the value of $\Ap$ based on historical measurements. From equation \ref{eq:const2chars} for $\Qerb$, and using the historical $\Bu = B_{u,0} = 4$ we arrive at $\Ap = A_{p,0} = \frac{1.0186}{Q_{\textrm{sim}}}$. This is consistent with the $b = 1.019 (\ERBf)$ from \cite{holdsworth1988implementing} and \cite{patterson1992complex}. Together with the historical $\Qerb = Q_{\textrm{sim}}$ derived from the same tuning curves used to obtain $B_{u,0} $ historically, we arrive at $A_{p,0} =  0.0252 \frac{4.37 \CF + 1}{\CF}$ where the CF is in kHz. This is equivalent to what we would have gotten had we estimated $\Ap$ and $\Bu$ simultaneously by fitting to the historical tuning curves.

\subsection{From $g_1$ and $Q_{\textrm{sim}}$}

The next estimate for $\Ap$ in figure \ref{fig:constEstVsCF} comes from $B_{u,1}$ and reported values of $\Qerb$ from simultaneous masking experiments, $\Qerb = Q_{\textrm{sim}} = \frac{10^3}{24.7} \frac{\CF}{4.37 \CF + 1}$ (with CF in kHz).  $Q_{\textrm{sim}}$ is reported based on tuning curves measured using simultaneous notched-noise masking psychoacoustic experiments performed in humans described in \cite{glasberg1990derivation}. This is the same set of experiments that were used to estimate the historical values for $n$ and $b$ of classical GTFs. Together with $\Bu = B_{u,1}$, and based on the equation for $\Qerb$ in equation \ref{eq:const2chars} (with $\bp = 1$), this results in an estimated  $\Ap = A_{p,1-1} = \frac{1.4334}{Q_{\textrm{sim}}} = 0.0354 \frac{4.37 \CF + 1}{CF}$ where CF is in kHz.

\subsection{From $g_1$ and $Q_{\textrm{forw}}$}

The final estimate of $\Ap$ in figure \ref{fig:constEstVsCF} comes from $B_{u,1}$ coupled with  $\Qerb = Q_{\textrm{forw}} = 11 \CF^{0.27}$ with CF in kHz is reported based on tuning curves measured using forward masking psychoacoustic experiments performed in humans for CFs between 1 and 8 kHz as described in \cite{oxenham2003estimates}. Together with $\Bu = B_{u,1}$, and based on the equation for $\Qerb$ in equation \ref{eq:const2chars} (with $\bp = 1$),  this results in an estimated $\Ap = A_{p,1-2} = \frac{1.4334}{Q_{\textrm{forw}}} = 0.1303 \CF^{-0.27}$ where CF is in kHz but $\Ap$ is dimensionless.


\subsection{Limitations}

In what follows, we note various caveats in our estimates and limitations of our approach for estimating filter constants from reported characteristics as suitable for auditory filterbanks mimicking healthy human processing. 

\subsubsection{Assumptions Regarding Constants}
We note that all estimates and constraints are valid only in the regime where the sharp-filter approximation remains accurate, i.e., small $\Ap$ (as is the case for humans) and when dealing with peak-centric characteristics. The estimates and constraints also build on our conclusion that $\Bu$ is CF-invariant. However, due to some conflicting observations described earlier, it may be useful to assess the validity of this conclusion.

\subsubsection{Distinguishing Classes}
In providing estimates for filter constants using reported values of characteristics, we were only concerned with estimates that hold across filter classes. We were not concerned with distinguishing which filter class is more appropriate. That would require an additional set of data and deriving additional expressions for distinguishing characteristics such as the asymmetry of the slope of the magnitude about the peak.

\subsubsection{Uncertainty in Characteristics}

Regardless of whichever set of filter characteristics we use, we must be cognizant of the error inherent in recordings and reported values - e.g. due to noisiness and assumptions used to obtain values of reported characteristics from measurements. In previous sections, we have considered everal cases were this may have an effect by studying the sensitivity of $\Bu$ estimates with respect to each of the characteristic ratios over the region of interest.

\subsubsection{Extrapolating from Different Species}

In the case of humans, we face an dearth of reliable and accepted reported values of filter characteristics, which leads us to integrating information from different species. In contrast, this would not be the case for cats, for instance, where revcor data can be used to even calculate the most peak-centric magnitude characteristics such as $\SB$. 

The $g_1$ used to estimate $\Bu$ is not only from a different species, but also depends on certain underlying assumptions from the Wiener Kernel methods. Ideally, when estimating filter constants for humans, we would have preferred to use readily available data from humans - specifically group delays from SFOAEs, $\Nsfoae$ along with $\Qerb$ from psychacoustic tuning curves. However, this approach requires accurate and precise values for $\eta$ to convert $\Nsfoae$ to $\NB$ which we do not have \footnote{If we do have this information, we must be sure to use $\Nsfoae$ and $\eta$ generated using the same experimental methods for obtaining SFOAEs}. This led us to using psychoacoustic $\Qerb$ and $g_1$ extrapolated from WKs from chinchilla ANFs.

In general regardless of the filter characteristics used, when estimating filter constants from characteristics for the case of humans, we may need to extrapolate characteristics that are expected to be species or CF-invariant from different species and locations and may even arise from different types of experiments (e.g. ANF, mechanical, or psychoacoustic), and to integrate information from different stimulus levels or those which are of different nature - e.g. from frequency responses and threshold tuning curves - this is also the case if we could use $\Nsfoae$ with $\Qerb$ to estimate the filter constants.

\subsubsection{Masking Paradigms}

One other caveat to keep in mind is the effect of choices made for the psychoacoustic experiments from which we obtain $\Qerb$. We used $\Qerb$ reported from psychoacoustic tuning curves using two specific forward and simultaneous masking paradigms as there is some debate regarding the appropriateness of various masker paradigms for estimating the frequency response magnitude. Additional experimental choices may also affect the reported values for $\Qerb$ from psychoacoustic experiments such as masker duration especially at lower CFs [\cite{lopez2024effect}].

\subsubsection{Alternative Methods}

Some of the advantages - and limitations, arise from our estimation of filter constants from filter characteristics. However, as previously mentioned, if we were solely interested in estimating one set of filter constants given a particular finely sampled and reliable tuning curve, then using our methods for estimating filter constants from filter characteristics is not the first choice. In this particular case, the most appropriate method  is to use the gold standard of simultaneously estimating filter constants by fitting the filter frequency response magnitude to the psychoacoustic tuning curves.

\section{Conclusions}

\subsection{Contributions}

In this paper, we designed human auditory filters using recently reported values of filter characteristics in a manner that facilitates understanding how variations in those characteristics influence the estimated filter constants. Towards our goal, we explored the full degrees of freedom of the filter rather than fixing the value of constants to those historically estimated based on simultaneous masking experiments in humans. The filter characteristics we included are primarily those that are peak-centric. Specifically, we analyzed the dependence of filter characteristics on filter constants using a sharp filter approximation for the transfer function of three classes of filters from the gammatone family. Our findings show that the choice of filter constants for auditory filters (e.g. the pole and filter exponent) greatly influence the behavior of auditory filters and ultimately the conclusions inferred from perceptual models as well as the performance of technologies that use auditory filterbanks. 

Indeed, it is our belief that fully utilizing the degrees of freedom of any given filter class is more important and allows us to access more behavior than switching over to another filter class - unless we are interested in filter behavior that cannot fundamentally be achieved by a certain class of filters - e.g. as relates to power spectrum asymmetry or nonlinear behavior. Hence, revisiting the parameter space was both theoretically and practically essential.

We also derived expressions for characteristic ratios that depend on either $\Bu$ or $\Ap$ including ratios such as $g$ for which we have reported values that we used to constrain the values of filter constants for a given species or applicatio. We studied the sensitivity of the filter characteristics to filter constants which informed which set of characteristics may be reliably used to design filters based on. We developed improved methods to design filters given specifications on sets of filter characteristics, and then demonstrated the accuracy of our findings regarding filter behavior (and implicitly our design methods) as applied to three different classes of realizable linear filters when $\Ap$ is small (as is the case for humans across CF). Lastly, we used these methods to design human auditory filterbanks by using the constraints on $\Bu$ from reported observations and characteristic ratios along with reported values of $\Qerb$ from both simultaneous and forward masking psychoacoustic experiments in humans.

\subsection{Future Directions}

Our expressions, constraints, and methods may further be used to directly and accurately design auditory filters if the native specifications provided are sets of filter characteristics. Additionally, these methods can be used to \textit{systematically} investigate the dependence of perceptual and technological study outcomes on filter characteristics – e.g. by studying the dependence on isolated characteristics (by varying ERB while fixing maximum group delay, or varying the 10 dB quality factor while fixing downward convexity). This is motivated by studies reporting sensitivity based on (usually ad hoc) variation of certain parameters – e.g. ERB and filter exponent. The outcomes influenced by changing filter characteristics include: intelligibility scores of bandpass-filtered speech [\cite{warren2004intelligibility}], accuracy of speech recognition [\cite{dimitriadis2010effects, slaney2014influence}], direction of arrival and sound source localization models [\cite{dietz2011auditory}], mutual information between articulatory gestures of vocal tracts and acoustic and perceptual features [\cite{ghosh2011processing}], and accuracy of speech intelligibility models for cochlear implants [\cite{cosentino2013cochlear}]. This work may also be used to understand the cochlea's role in perception via underlying unified models [\cite{alkhairy2019analytic}].

We note that our interest was not only to estimate the filter constants given a particular tuning curve - which lead to the filter characteristics-based approach. However, we encourage fitting the filters to the forward masking tuning curves and estimating the filter constants  to check our estimates for humans based on these data. It is also useful to then compute $g$ from the filter constants estimated in this way in order to check the validity of our extrapolation of $g$ across species and CFs.

Our analyses and methods may be extended to include specifications on combined spectrotemporal characteristics as is relevant for studying certain perceptual functions [\cite{moore2008role}]. Future work may include other investigation and derivation of methods including filter characteristics specific to certain filter classes (such as asymmetry), and using those characteristics for estimating filter constants. Measure of uncertainty may also be derived for the estimated filter constants based on our expressions for filter characteristics and can be used if we are provided with the standard deviation of reported characteristics.

We have tested the accuracy of our results and methods for three related classes of auditory filters from the gammatone family of filters and expect that our conclusions hold for other such classes in the same family as well, but it is appropriate to test for accuracy for those classes of filters if we are interested in designing filters of those classes - particularly if their shape is fundamentally different in the peak region or they have additional degrees of freedom that affect behavior in the peak region. Lastly, it is desirable to build on this work towards quasilinear filters and  handling nonlinear versions of the filters as is especially relevant for hearing loss simulations [\cite{irino2020gammachirp}].

\appendix

\section{Appendix: Derivation of Expression for ERB in Terms of Filter Constants}

Here we derive our expressions for $\ERBB$ and its associated quality factor (in equation \ref{eq:const2chars}) in terms of the filter constants. The $\ERBB$ (which we will refer to as ERB in this section for simplicity) is defined as follows for a filter with transfer function $H$:
\begin{equation}
    \textrm{ERB} \triangleq \frac{1}{|H(\Bcenter)|^2} \defint{0}{\infty}{|H(\B)|^2}{\B} \;,
\end{equation}

where $H$ may be $\HP, \HV$, $\HGTF$, or the transfer function of any other bandpass filter. We note that when computing the ERB numerically, the lower limit of integration is replaced with a `small-enough' value $\B_1$ and the upper limit of integration is replaced with a `large-enough' value $\B_2$.

Let us return to the definition of ERB in the above equation, substitute $H$ by $G = \HSharp$ which approximates the transfer function of realizable filters of interest. This leads to,

\begin{equation}
    \textrm{ERB} \approx \frac{1}{|G(\Bcenter)|^2} \defint{0}{\infty}{|G(\B)|^2}{\B} \;,
\end{equation}

or equivalently,

\begin{equation}
\text{ERB}  \approx \frac{1}{|G_{max}|^2} I_\infty \;, 
\label{eq:ERBfromGmaxAndIinfty}
\end{equation}

with,

\begin{equation}
    I_\infty  = \defint{-\infty}{\infty}{|G(\B)|^2}{\B} \;,
\end{equation}

where we have replaced the lower limit of integration by $-\infty$ for simplicity. This is appropriate because $|G|$ (the magnitude of $\HSharp$) is negligible for $\B < 0$ in the parameter region of interest.

From the expression for $G$ (equation \ref{eq:Hsharp} and its magnitude - see \cite{alkhairy2024characteristics}), we derive, 

\begin{equation}
    |G_{max}|^2  = \Ap^{-2\Bu}  \;.
    \label{eq:Gmax2}
\end{equation}

To derive our expression for $I_\infty$, we rewrite it as follows,
\small
\begin{equation}
    I_\infty  = \lim_{n \xrightarrow{} \infty} J(\B) \bigg|_{-n}^{n} \;,
    \label{eq:IinfEq}
\end{equation}
\normalsize

with,

\small
\begin{equation}
    \begin{aligned}
    J(\B) & = \defint{}{\B}{|G(\B')|^2}{\B'} \\
    & = \defint{}{\B}{\bigg( (\Ap^2 + (\B'-\bp)^2 )^{-\Bu/2} \bigg)^2}{\B'} \quad \\
    & = \underbrace{  \frac{\B - \bp}{\Ap^{2\Bu}}  }_{\triangleq \zeta} \ _2F_1 (\frac{1}{2}, \Bu; \frac{3}{2}; - (\frac{\B-\bp}{\Ap})^2) \;,
\end{aligned}
\label{eq:J}
\end{equation}
\normalsize

This expression is in terms of the Gauss Hypergeometric Function (GHF), which - for the case of $|z|>1$ \footnote{which is the relevant form of the GHF in our case taking $\B \xrightarrow{} \infty, -\infty$}, is expressed as,

\small
\begin{multline}
    _2F_1(a,b;c;z)  = \frac{\Gamma(b-a) \Gamma(c)}{\Gamma(b) \Gamma(c-a)} (-z^{-1})^{a} \ _2F_1(a,a-c+1;a-b+1;z^{-1}) \\
    + \frac{\Gamma(a-b) \Gamma(c)}{\Gamma(a) \Gamma(c-b)} (-z^{-1})^{b} \ _2F_1(b, b-c+1; -a+b+1; z^{-1})   \;.
\label{eq:GHFlargez}
\end{multline}

\normalsize
To use \removeInShortVer{equation }(\ref{eq:GHFlargez}) for \removeInShortVer{equation }(\ref{eq:J}) of the ERB, we define $z$ and $y$ as
\begin{equation}
    z = -y = - (\frac{\B-\bp}{\Ap})^2 \;,
\end{equation}

with $|\B| \xrightarrow{} \infty \implies z \xrightarrow{} - \infty$, and with $a= \frac{1}{2} , b = \Bu, c= \frac{3}{2}$, and $\Gamma(1/2) = \sqrt{\pi}, \Gamma(3/2) = \sqrt{\pi}/2, \Gamma(1) = 1$.

As a result, we have,
\small
\begin{multline}
    _2F_1 (\frac{1}{2}, \Bu; \frac{3}{2}; -y) = \frac{\Gamma(\Bu-\frac{1}{2}) \frac{\sqrt{\pi}}{2} }{\Gamma(\Bu) 1} y^{-\frac{1}{2}} \ _2F_1(\frac{1}{2}, 0; \frac{3}{2} -\Bu; -y^{-1})\\
    + \frac{\Gamma(\frac{1}{2}-\Bu) \frac{\sqrt{\pi}}{2}}{\sqrt{\pi} \Gamma(\frac{3}{2}-\Bu)} y^{-\Bu} \ _2F_1(\Bu, \Bu - \frac{ 1}{2}; \Bu + \frac{1}{2}; -y^{-1})
\end{multline}
\normalsize

The first GHF in the above expression is simply $=1$ due to the zero argument. We expand the second term using the formula for GHF for $|z|<1$,

\begin{equation}
    _2F_1(a,b;c;z) = 1 + \frac{ab}{c} z + O(z^2)
    \label{eq:GHFsmallzseries}
\end{equation}

and obtain,

\small
\begin{multline}
    _2F_1 (\frac{1}{2}, \Bu; \frac{3}{2}; -y) = a_1 y^{-\frac{1}{2}} \\ 
    + a_2 y^{-\Bu} \big( 1 + \Bu \frac{\Bu - \frac{1}{2}}{\Bu + \frac{1}{2}} (-y^{-1}) + O(y^{-2}) \big) \;,
\end{multline}
\normalsize

with,
\small
\begin{equation}
    \begin{aligned}
        a_1 & = \frac{\sqrt{\pi}}{2} \frac{\Gamma(\Bu-\frac{1}{2})  }{\Gamma(\Bu)}\\
        a_2 & = \frac{1}{2} \frac{\Gamma(\frac{1}{2} -\Bu ) }{ \Gamma(\frac{3}{2} -\Bu)} \;.
    \end{aligned}
\end{equation}
\normalsize

Substituting the above expansion for $_2F_1 (\frac{1}{2}, \Bu; \frac{3}{2}; -y)$ for large $y$ into \removeInShortVer{equation }(\ref{eq:J}) for $J(\B)$, we get,

\begin{multline}
    J(\B) = \frac{\B - \bp}{\Ap^{2\Bu}}  a_1 y^{-\frac{1}{2}} \\
    + a_2 y^{-\Bu} \big( 1 + \Bu \frac{\Bu - \frac{1}{2}}{\Bu + \frac{1}{2}} (-y^{-1}) + O(y^{-2}) \big) \;.
\end{multline}

As $|\B| \xrightarrow{} \infty$, all but the first term go to zero due to $\Bu > 1$. This results in,

\small
\begin{equation}
\begin{aligned}
    \lim_{|\B| \xrightarrow{} \infty} J(\B) & = \frac{\B - \bp}{\Ap^{2\Bu}}  a_1 y^{-\frac{1}{2}} \\
    & = \frac{\B - \bp}{\Ap^{2\Bu}}  a_1 ((\frac{\B - \bp}{\Ap})^2)^{-\frac{1}{2}} \\
    & =  \frac{\B - \bp}{\Ap^{2\Bu}}  a_1 \frac{\Ap}{|\B - \bp|} \\
    & = a_1 \Ap^{1-2\Bu} \frac{\B - \bp}{|\B - \bp|} \;.
\end{aligned}
\end{equation}
\normalsize

Therefore, 
\begin{equation}
         \lim_{\B \xrightarrow{} \infty} J(\B) = a_1 \Ap^{1-2\Bu} \;,
\end{equation}
and,
\begin{equation}     
         \lim_{\B \xrightarrow{} -\infty} J(\B) = -a_1 \Ap^{1-2\Bu} \;,
\end{equation}
which - by \removeInShortVer{equation }(\ref{eq:IinfEq}), result in,
\begin{equation}     
    I_\infty = 2a_1 \Ap^{1-2\Bu} = \sqrt{\pi} \frac{\Gamma (\Bu - \frac{1}{2})}{\Gamma(\Bu)}\Ap^{1-2\Bu} \;.
\end{equation}

Consequently based on equation \ref{eq:ERBfromGmaxAndIinfty} and using equation \ref{eq:Gmax2}, we arrive at,

\small
\begin{equation}   
    \text{ERB}_\infty = \sqrt{\pi} \frac{\Gamma (\Bu - \frac{1}{2})}{\Gamma(\Bu)}\Ap \;,
    \label{eq:ERBgamma}
\end{equation}
\normalsize

and its associated quality factor,
\begin{equation}     
         Q_{erb} = \frac{\bp}{\sqrt{\pi}\Ap} \frac{\Gamma (\Bu)}{\Gamma(\Bu - \frac{1}{2})} \;,
\end{equation}

as shown in equation \ref{eq:const2chars}. 

We note that for positive integer values of $\Bu$ (and $\bp = 1$), the $\Qerb$ we derived above based on $\HSharp$ is very well approximated by the previously derived expression for GTFs, which we denote in the following equation as  $Q_{erb, \text{GTF}}$. The expression for $Q_{erb, \text{GTF}}$ was found by deriving the $\ERBf$ from the impulse response of the classical GTFs - by using Parseval's theorem as described in \cite{holdsworth1988implementing} and detailed in \cite{darling1991properties}.

\begin{equation}
\begin{aligned}
    Q_{erb, \text{GTF}} &= \frac{\CF}{\ERBf} \\
    & = \frac{\CF}{\pi b}\frac{(n-1)!^2 2^{2n-2}}{(2n - 2)!} \\
    & = \frac{1}{\pi \Ap} \frac{(n-1)!^2 2^{2n-2}}{(2n-2)!} \;.
\end{aligned}
\end{equation}




\bibliographystyle{abbrvnat} 

\bibliography{references}

\end{document}

%% file: samiyacommands.tex
\usepackage{float}

\usepackage{mathtools, cuted}

\usepackage{tabularx}
\usepackage{amsmath}
\usepackage{amssymb}
\usepackage{mathalfa}

\usepackage{nccmath}

\newcommand{\Qn}{Q_n}
\newcommand{\SB}{S_{\beta}}
\newcommand{\Qerb}{Q_{\textrm{erb}}}
\newcommand{\Nsfoae}{N_{\textrm{sfoae}}}
\newcommand{\ERBB}{\textrm{ERB}_\B}
\newcommand{\ERBf}{\textrm{ERB}_f}

\newcommand{\defint}[4]{\int_{#1}^{#2}{#3} d{#4}}

\newcommand{\B}{\beta}

\newcommand{\expn}[1]{e^{#1}}

\newcommand{\Ap}{A_p}
\newcommand{\bp}{b_p}
\newcommand{\Bu}{B_u}

\newcommand{\pconj}{\overline{p}}

\renewcommand{\xi}{x_i}

\newcommand{\level}[1]{\text{mag}\{{#1}\}}

\newcommand{\phase}[1]{\text{phase}\{{#1}\}}

\newcommand{\ttilde}{\tilde{t}}

\renewcommand{\Im}{\text{Im}}

\usepackage{lipsum}

\usepackage{longtable} 

\newcommand{\HP}{H_P}
\newcommand{\HSharp}{H_\textrm{sharp}}
\newcommand{\HV}{H_V}

\newcommand{\HGTF}{H_{GTF}}

\usepackage{eqnarray,amsmath}

\newcommand{\appropto}{\mathrel{\vcenter{
  \offinterlineskip\halign{\hfil$##$\cr
    \propto\cr\noalign{\kern2pt}\sim\cr\noalign{\kern-2pt}}}}}

\newcommand{\Bcenter}{\B_{peak}}
\newcommand{\phiaccum}{\phi_{accum}}
\newcommand{\BWndBbeta}{\text{BW}_{n,\B}}

\newcommand{\CF}{\text{CF}}

\usepackage{diagbox}

\newcommand{\hGTF}{h_{GTF}}

\newcommand{\HPsharp}{\HSharp}

\usepackage{caption}
\usepackage{subcaption}

\newcommand{\NB}{N_\B}

%% file: references.bib
@article{warren2004intelligibility,
  title={Intelligibility of bandpass filtered speech: Steepness of slopes required to eliminate transition band contributions},
  author={Warren, Richard M and Bashford Jr, James A and Lenz, Peter W},
  journal={The Journal of the Acoustical Society of America},
  volume={115},
  number={3},
  pages={1292--1295},
  year={2004},
  publisher={Acoustical Society of America}
}

@article{dimitriadis2010effects,
  title={On the effects of filterbank design and energy computation on robust speech recognition},
  author={Dimitriadis, Dimitrios and Maragos, Petros and Potamianos, Alexandros},
  journal={IEEE transactions on audio, speech, and language processing},
  volume={19},
  number={6},
  pages={1504--1516},
  year={2010},
  publisher={IEEE}
}

@inproceedings{slaney2014influence,
  title={The influence of pitch and noise on the discriminability of filterbank features.},
  author={Slaney, Malcolm and Seltzer, Michael L},
  booktitle={INTERSPEECH},
  volume={14},
  pages={2263--2267},
  year={2014}
}

@article{dietz2011auditory,
  title={Auditory model based direction estimation of concurrent speakers from binaural signals},
  author={Dietz, Mathias and Ewert, Stephan D and Hohmann, Volker},
  journal={Speech Communication},
  volume={53},
  number={5},
  pages={592--605},
  year={2011},
  publisher={Elsevier}
}

@article{ghosh2011processing,
  title={Processing speech signal using auditory-like filterbank provides least uncertainty about articulatory gestures},
  author={Ghosh, Prasanta Kumar and Goldstein, Louis M and Narayanan, Shrikanth S},
  journal={The Journal of the Acoustical Society of America},
  volume={129},
  number={6},
  pages={4014--4022},
  year={2011},
  publisher={AIP Publishing}
}

@article{cosentino2013cochlear,
  title={Cochlear implant filterbank design and optimization: A simulation study},
  author={Cosentino, Stefano and Falk, Tiago H and McAlpine, David and Marquardt, Torsten},
  journal={IEEE/ACM Transactions on Audio, Speech, and Language Processing},
  volume={22},
  number={2},
  pages={347--353},
  year={2013},
  publisher={IEEE}
}

@article{moore2008role,
  title={The role of temporal fine structure processing in pitch perception, masking, and speech perception for normal-hearing and hearing-impaired people},
  author={Moore, Brian CJ},
  journal={Journal of the Association for Research in Otolaryngology},
  volume={9},
  number={4},
  pages={399--406},
  year={2008},
  publisher={Springer}
}

@article{irino2020gammachirp,
  title={The gammachirp auditory filter and its application to speech perception},
  author={Irino, Toshio and Patterson, Roy D},
  journal={Acoustical Science and Technology},
  volume={41},
  number={1},
  pages={99--107},
  year={2020},
  publisher={ACOUSTICAL SOCIETY OF JAPAN}
}

@article{bentsen2011human,
  title={Human cochlear tuning estimates from stimulus-frequency otoacoustic emissions},
  author={Bentsen, Thomas and Harte, James M and Dau, Torsten},
  journal={The Journal of the Acoustical Society of America},
  volume={129},
  number={6},
  pages={3797--3807},
  year={2011},
  publisher={AIP Publishing}
}

@misc{Guinan,
  author       = {Guinan Jr, J. J.},
  howpublished = {Personal communication},
  year         = {2017},
}

@article{lineton2009comparing,
  title={Comparing two proposed measures of cochlear mechanical filter bandwidth based on stimulus frequency otoacoustic emissions},
  author={Lineton, Ben and Wildgoose, Catriona},
  journal={The Journal of the Acoustical Society of America},
  volume={125},
  number={3},
  pages={1558--1566},
  year={2009},
  publisher={AIP Publishing}
}

@article{schairer2006use,
  title={Use of stimulus-frequency otoacoustic emission latency and level to investigate cochlear mechanics in human ears},
  author={Schairer, Kim S and Ellison, John C and Fitzpatrick, Denis and Keefe, Douglas H},
  journal={The Journal of the Acoustical Society of America},
  volume={120},
  number={2},
  pages={901--914},
  year={2006},
  publisher={AIP Publishing}
}

@article{valero2012gammatone,
  title={Gammatone cepstral coefficients: Biologically inspired features for non-speech audio classification},
  author={Valero, Xavier and Alias, Francesc},
  journal={IEEE transactions on multimedia},
  volume={14},
  number={6},
  pages={1684--1689},
  year={2012},
  publisher={IEEE}
}

@inproceedings{jin2017application,
  title={Application of Gammatone filter bank to active noise control algorithm},
  author={Jin, Yan and Su, Haitao and Xu, Cuifeng and Guo, Qing},
  booktitle={2017 IEEE International Conference on Signal Processing, Communications and Computing (ICSPCC)},
  pages={1--5},
  year={2017},
  organization={IEEE}
}

@article{matsumoto2011application,
  title={Application of an auditory filter for the evaluation of sounds and sound fields},
  author={Matsumoto, Yuki and Suzuki, Masahiro and Ogushi, Hisako and Omoto, Akira},
  journal={Building Acoustics},
  volume={18},
  number={1-2},
  pages={175--188},
  year={2011},
  publisher={SAGE Publications Sage UK: London, England}
}

@article{glasberg1990derivation,
  title={Derivation of auditory filter shapes from notched-noise data},
  author={Glasberg, Brian R and Moore, Brian CJ},
  journal={Hearing research},
  volume={47},
  number={1-2},
  pages={103--138},
  year={1990},
  publisher={Elsevier}
}

@incollection{patterson1992complex,
  title={Complex sounds and auditory images},
  author={Patterson, Roy D and Robinson, KEN and Holdsworth, John and McKeown, Denis and Zhang, C and Allerhand, Michael},
  booktitle={Auditory physiology and perception},
  pages={429--446},
  year={1992},
  publisher={Elsevier}
}

@article{lopez2024effect,
  title={Effect of stimulus duration on estimates of human cochlear tuning},
  author={Lopez-Ramos, David and Eustaquio-Martin, Almudena and Lopez-Bascuas, Luis E and Lopez-Poveda, Enrique A},
  journal={Hearing Research},
  volume={451},
  pages={109080},
  year={2024},
  publisher={Elsevier}
}

@article{oxenham2003estimates,
  title={Estimates of human cochlear tuning at low levels using forward and simultaneous masking},
  author={Oxenham, Andrew J and Shera, Christopher A},
  journal={Journal of the Association for Research in Otolaryngology},
  volume={4},
  number={4},
  pages={541--554},
  year={2003},
  publisher={Springer}
}

@article{holdsworth1988implementing,
  title={Implementing a gammatone filter bank},
  author={Holdsworth, John and Nimmo-Smith, Ian and Patterson, Roy and Rice, Peter},
  journal={Annex C of the SVOS Final Report: Part A: The Auditory Filterbank},
  volume={1},
  pages={1--5},
  year={1988}
}

@article{slaney1993efficient,
  title={An efficient implementation of the Patterson-Holdsworth auditory filter bank},
  author={Slaney, Malcolm and others},
  journal={Apple Computer, Perception Group, Tech. Rep},
  volume={35},
  number={8},
  pages={1--42},
  year={1993},
  publisher={Citeseer}
}

@article{katsiamis2007practical,
  title={Practical gammatone-like filters for auditory processing},
  author={Katsiamis, Andreas G and Drakakis, Emmanuel M and Lyon, Richard F},
  journal={EURASIP Journal on Audio, Speech, and Music Processing},
  volume={2007},
  number={1},
  pages={063685},
  year={2007},
  publisher={Springer}
}

@article{shera2010otoacoustic,
  title={Otoacoustic estimation of cochlear tuning: validation in the chinchilla},
  author={Shera, Christopher A and Guinan Jr, John J and Oxenham, Andrew J},
  journal={Journal of the Association for Research in Otolaryngology},
  volume={11},
  number={3},
  pages={343--365},
  year={2010},
  publisher={Springer}
}

@article{shera2003stimulus,
  title={Stimulus-frequency-emission group delay: A test of coherent reflection filtering and a window on cochlear tuning},
  author={Shera, Christopher A and Guinan Jr, John J},
  journal={The Journal of the Acoustical Society of America},
  volume={113},
  number={5},
  pages={2762--2772},
  year={2003},
  publisher={Acoustical Society of America}
}

@article{alkhairy2024characteristics,
  title={Characteristics-based design of generalized-exponent bandpass filters},
  author={Alkhairy, Samiya A},
  journal={IEEE Access},
  year={2025},
  publisher={IEEE}
}

@article{alkhairy2019analytic,
  title={An analytic physically motivated model of the mammalian cochlea},
  author={Alkhairy, Samiya A and Shera, Christopher A},
  journal={The Journal of the Acoustical Society of America},
  volume={145},
  number={1},
  pages={45--60},
  year={2019},
  publisher={AIP Publishing}
}

@inproceedings{alkhairy2024cochlear,
  title={Cochlear wave propagation and dynamics in the human base and apex: Model-based estimates from noninvasive measurements},
  author={Alkhairy, Samiya A},
  booktitle={AIP Conference Proceedings},
  volume={3062},
  number={1},
  pages={020010},
  year={2024},
  organization={AIP Publishing LLC}
}

@article{alkhairy2025rational,
  title={Rational-Exponent Filters with Applications to Generalized Exponent Filters},
  author={Alkhairy, Samiya A},
  journal={IEEE Transactions on Circuits and Systems I: Regular Papers},
  year={2025},
  publisher={IEEE}
}

@article{tan2003phenomenological,
  title={A phenomenological model for the responses of auditory-nerve fibers. II. Nonlinear tuning with a frequency glide},
  author={Tan, Qing and Carney, Laurel H},
  journal={The Journal of the Acoustical Society of America},
  volume={114},
  number={4},
  pages={2007--2020},
  year={2003},
  publisher={Acoustical Society of America}
}

@article{robles2001mechanics,
  title={Mechanics of the mammalian cochlea},
  author={Robles, Luis and Ruggero, Mario A},
  journal={Physiological reviews},
  volume={81},
  number={3},
  pages={1305--1352},
  year={2001},
  publisher={American Physiological Society Bethesda, MD}
}

@article{darling1991properties,
  title={Properties and implementation of the gammatone filter: a tutorial},
  author={Darling, AM},
  journal={Speech Hearing and Language, Work in Progress, University College London, Department of Phonetics and Linguistics},
  pages={43--61},
  year={1991}
}
